\documentclass[reprint, NumberedRefs]{JASA}

 \usepackage{amsmath,amssymb,amsfonts}
\usepackage{algorithmicx}
\usepackage{array}
\usepackage[caption=false,font=footnotesize]{subfig}
\usepackage{url}
\usepackage{stmaryrd}
\usepackage{tcolorbox}
\usepackage{graphicx,graphics,color,psfrag,multirow}
\usepackage{subfloat}
\usepackage{url}

\allowdisplaybreaks[4]

\newcommand{\bc}[1]{\mbox{\boldmath $\mathcal{#1}$}}

\newcommand{\bs}[1]{\boldsymbol{#1}}
\newcommand{\F}{\mathrm{F}}
\newcommand{\T}{\mathrm{T}}

\usepackage{diagbox}

\begin{document}

\title[JASA/Sample JASA Article]{Striking The Right Balance:  Three-Dimensional Ocean Sound Speed Field Reconstruction Using Tensor Neural Networks}

\author{Siyuan Li, Lei Cheng*,}
\thanks{* denotes corresponding author (lei\_cheng@zju.edu.cn)}
\author{Ting Zhang, Hangfang Zhao and Jianlong Li}

\affiliation{College of Information Science and Electronic Engineering, Zhejiang University, Hangzhou, 310027, China}

\preprint{Author, JASA}	

\date{\today}

\begin{abstract}

Accurately reconstructing a three-dimensional ocean sound speed field (3D SSF) is essential for various ocean acoustic applications, but the sparsity and uncertainty of sound speed samples across a vast ocean region make it a challenging task. To tackle this challenge,  a large body of reconstruction methods has been developed, including spline interpolation, matrix/tensor-based completion, and deep neural networks-based reconstruction. However, a principled analysis of their effectiveness in 3D SSF reconstruction is still lacking. This paper performs a thorough analysis of the reconstruction error and highlights the need for a balanced representation model that integrates both expressiveness and conciseness. To meet this requirement, a 3D SSF-tailored tensor deep neural network is proposed, which utilizes tensor computations and deep neural network architectures to achieve remarkable 3D SSF reconstruction. The proposed model not only includes the previous tensor-based SSF representation model as a special case, but also has a natural ability to reject noise. The numerical results using the South China Sea 3D SSF data demonstrate that the proposed method outperforms state-of-the-art methods. The code is available at https://github.com/OceanSTARLab/Tensor-Neural-Network.

\end{abstract}


\maketitle


\section{\label{sec:1} Introduction}

\begin{figure*}[t]
    \center
    \includegraphics[width=2\reprintcolumnwidth]{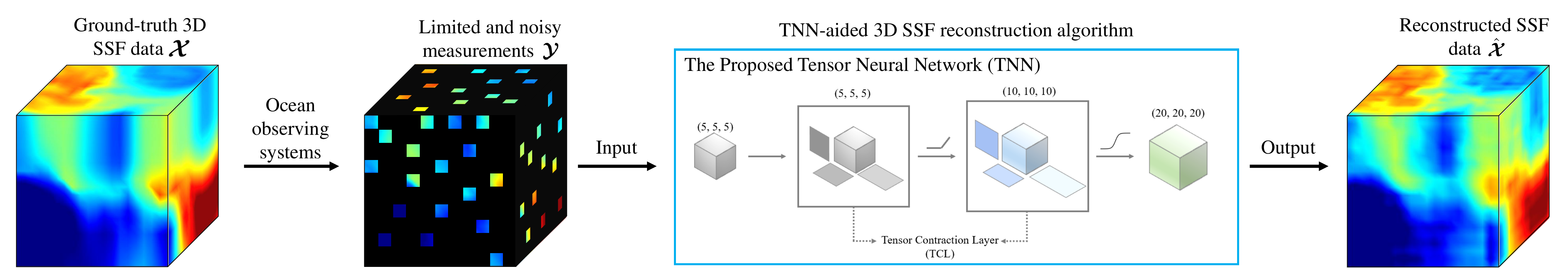}
    \caption{ Illustration of the reconstruction system considered in this paper, with the proposed 3D SSF-tailored tensor deep neural network serving as the representation model.}
    \label{fig:framework}
    \hrule
\end{figure*}

Three-dimensional (3D) ocean sound speed fields (SSFs), which characterize sound propagation over 3D geographical regions \cite{jensen2011computational}, form the stepping stone towards the success of a myriad of ocean acoustic applications, including underwater detection,\cite{li2011time} localization,\cite{michalopoulou2021matched} and communications.\cite{qu2014two} To realize sound speed awareness, ocean observing systems have been rapidly developing in recent years, leading to the proliferation of intelligent underwater floats \cite{nystuen2011interpreted} and vehicles.\cite{chen2018experimental} Despite these advancements, sound speed samples are still sparsely scattered across a vast ocean region (see Fig.~\ref{fig:framework}), making crafting an accurate and fine-grained 3D SSF a considerable challenge.

The crux of the matter is that the number of samples is much less than the unknowns,  resulting in a highly under-determined inverse problem.\cite{tarantola2005inverse} 
To make the problem tractable, the primary approach involves modeling the prior knowledge of the unknowns and incorporating that information into the reconstruction process. 
This has led to a sizable body of related works on reconstruction methods,\cite{li2023graph,saragadam2022deeptensor,xu2021overfitting,kanatsoulis2018hyperspectral,shrestha2022deep} not limited to SSF reconstruction, that have a near-universal three-step approach: 1) proposing a representation model that incorporates prior information, 2) learning the model parameters from limited samples, and 3) reconstructing the underlying signal using the learned model.

For instance, the simple yet widely-adopted spline interpolation method assumes that the underlying signal is a linear combination of smooth Green functions centered at the samples, \cite{sandwell1987biharmonic} which correspond to the linear regression model in machine learning (see detailed discussion in Sec. II of Ref.~\citen{li2023graph}). Using linear regression as a representation model, the spline interpolation method starts by learning the optimal weights for the Green functions from the samples, and then uses the learned model to reconstruct the unknowns.  A step forward is the Gaussian process regression (GPR),\cite{rasmussen2004gaussian} which models the underlying signal as a Gaussian random process with a kernel function describing spatial correlations. GPR first learns the hyper-parameters of the model from the samples and then reconstructs the unobserved values. In recent years, matrix/tensor decompositions and deep neural networks have been increasingly used as representation models for various data completion tasks, leading to state-of-the-art (SOTA) results. Our task is closely related to recent advances in 3D data inverse problems and field reconstruction, which includes video/image inpainting \cite{long2021bayesian,xu2021overfitting}, hyperspectral reconstruction \cite{kanatsoulis2018hyperspectral,ding2020hyperspectral}, and radio map cartography \cite{shrestha2022deep,zhang2020spectrum}.

The burgeoning literature on reconstruction methods invites the question: {\it Among numerous ways, which one is the most effective for 3D ocean SSF reconstruction?} Rather than exhaustively comparing various methods, this paper sets out to answer this question in a principled manner by delving into the analysis of reconstruction error. Specifically, inspired by the well-known bias-variance trade-off in machine learning,\cite{bishop2006pattern} the reconstruction error can be mainly decomposed into two parts: the {\it representation error}, which measures the model's capacity to fit the underlying signal (e.g., 3D SSF), and the {\it identification error}, which assesses whether the desired model can be uniquely learned from limited samples.

To clarify this notion, consider the case where an over-parameterized deep neural network (i.e., one with more parameters than the sound speed samples) is used as the representation model.  Its high expressive power \cite{hornik1989multilayer} makes it likely to precisely represent the underlying 3D SSF, resulting in negligible representation error. However, insufficient samples and numerous unknown model parameters can lead to difficulties in identifying the optimal parameter configuration across a vast solution space. Unless strong regularizations (e.g., early stopping \cite{yao2007early}) are carefully employed, the overwhelming identification error may lead to poor reconstruction. On the other hand, choosing a simpler representation model with fewer parameters makes it easier to uniquely learn the model parameters from limited samples, but may result in a higher representation error.

The reconstruction error analysis above {\it highlights the need for a balanced representation model} in 3D SSF reconstruction, with both high expressive power to reduce representation error and conciseness (in terms of both parameter numbers and mathematical operations) to minimize identification error. To meet this requirement, we propose a 3D SSF-tailored tensor deep neural network, drawing on the advantages of both recent {\it succinct tensor-based} ocean SSF representation \cite{chen2022tensor,cheng2022tensor} and {\it powerful deep neural network-based} image representation.\cite{ulyanov2018deep} The network, shown in Fig.~\ref{fig:framework}, aims to provide a concise and accurate representation of SSFs.

The proposed 3D SSF-tailored tensor deep neural network leverages the strengths of tensor computations and deep neural network architectures to achieve remarkable 3D SSF reconstruction. The tensor computations,  which are with a few parameters and serve as the model's backbone,  effectively exploit the multi-dimensional correlations among sound speeds.\cite{chen2022tensor,cheng2022tensor} On the other hand, the enrollment of deep neural network architectures enhances the model's expressive power, especially in capturing fine-grained variations of sound speeds. Due to its conciseness, a simple gradient descent algorithm \cite{nesterov2003introductory} is sufficient to identify the parameters that yield a quite good reconstruction from limited samples. Relying on its high expressive power, the learned model successfully reconstructs 3D SSFs with intricate details.  Two key features of the proposed method are highlighted in this paper. First, the model has a natural ability to reject noise. This property is backed by theoretical proof under the one-layer assumption, and experimental verification in multiple layer scenarios. Second, the proposed model includes the tensor Tucker decomposition model as a specific case, which is strongly linked to classical SSF basis functions such as empirical orthogonal functions (EOFs).\cite{cheng2022tensor}

Moreover, the proposed model can easily incorporate additional structural assumptions to boost the reconstruction performance further. As an example, we integrate total variation\cite{rudin1992nonlinear} regularizations in this paper to exploit the spatial smoothness of SSF. Finally, to corroborate the advantages of the proposed model, we conduct comprehensive numerical experiments using real-life 3D SSF data and compare the performance of the proposed approach with SOTA reconstruction methods. The encouraging results of the proposed algorithm not only affirm our theoretical analysis, but also demonstrate the effectiveness of viewing the reconstruction task from the reconstruction error analysis perspective.

{\it Notation}: Lower- and uppercase bold letters (e.g., $\mathbf{x}$ and $\mathbf{X}$) are used to denote the vectors and matrices. Higher-order tensors (order three or higher) are denoted by upper-case bold calligraphic letters. For a tensor $\bc{X}$, $\mathbf{X}_{(p)}$ stands for its model-$p$ unfolding matrix. $\bc{X}\times_p\mathbf{A}$ denotes the $p$-mode product between tensor $\bc{X}$ and matrix $\mathbf{A}$. The Kronecker product is denoted by $\otimes$. The Hadamard product is denoted by $*$. $\|\cdot\|_\F$ stands for the Frobenius norm. $\langle\cdot,\cdot\rangle$ denotes the tensor inner product. The superscript $^\T$ stands for transposition. $\mathbb{R}$ and $\mathbb{C}$ are the field of real and complex numbers, respectively.

\section{\label{sec:2} Reconstruction and Error Analysis}
In this section, we formulate the ocean 3D SSF reconstruction task as an optimization problem, and present a unified framework to analyze the reconstruction error. Based on this framework, two widely used reconstruction methods are analyzed. 

\subsection{Reconstruction Problem Formulation}

The reconstruction system considered in this paper is illustrated in Fig.~\ref{fig:framework}.  Assuming that the sound speed samples are contaminated by independently and identically distributed (i.i.d) Gaussian noise, the SSF reconstruction problem can be formulated as
\begin{align}
    \min_{\bc{X}}\|\bc{Y}-\bc{O}* \bc{X}\|_\F^2,
    \label{eq1}
\end{align}
where $\bc{X}\in \mathbb{R}^{I\times J\times K}$ is the ground-truth 3D SSF (to be reconstructed) and $\bc{O}$ is a binary indicator tensor with $\bc{O}_{i,j,k}=1$ if entry $(i,j,k)$ is observed. Observation tensor $\bc{Y} \in \mathbb{R}^{I\times J\times K}$ collects noisy and limited SSF samples, i.e., $\bc{Y}_{i,j,k}$ is the sampled sound speed value if $\bc{O}_{i,j,k}=1$, and otherwise equals to zero.

Denote $N$ as the number of observed (non-zero) entries in $\bc Y$,  and $T = IJK$ as the total number of entries in $\bc{X}$. The SSF reconstruction problem in Eq.~\eqref{eq1} is under-determined because $N$ is usually much smaller than $T$. To tackle this challenge, existing methods (e.g., Refs.~\citen{li2014spatial,sandwell1987biharmonic,li2023graph,ulyanov2018deep,liu2012tensor}) usually take the following three steps.

{\it1) SSF Representation:} Since ocean sound speeds exhibit strong spatial correlations, SSF can be effectively represented by one model with only {\it a few} parameters, as demonstrated in prior literature.\cite{bianco2017dictionary,cheng2022tensor}
Mathematically, assume that ocean SSF is represented by:
\begin{align}
    \bc{X} \approx \bc D( \bs \Theta),
\end{align}
where $\bc D(\cdot)$ denotes an SSF representation model and $\bs \Theta$ is the set of model parameters.

{\it 2) Parameter Learning:}  Using the representation model $\bc D(\cdot)$, the SSF reconstruction problem defined in Eq.~\eqref{eq1} can be recast as
\begin{align}
    \min_{\bs \Theta}\|\bc{Y}-\bc{O}* \bc D( \bs \Theta)\|_\F^2.
    \label{eq:rec}
\end{align}
The model parameter estimate  $\hat{\bs \Theta}$ is then obtained by solving problem~\eqref{eq:rec}.

{\it 3) SSF Reconstruction:}  Using the learned model parameter  $\hat{\bs \Theta}$,  the SSF is reconstructed by $\hat{\bc{X}}=\bc D(\hat{\bs \Theta})$.

\subsection{Reconstruction Error Analysis}

We proceed to conduct a theoretical analysis of the reconstruction error, which is defined as $E=\|\hat{\bc{X}}-\bc{X}\|_\F^2$. 
Since the reconstruction error is a least-squares function, its associated error analysis, such as the bias-variance trade-off, has already been developed in the machine learning and signal processing literature. For example, refer to Page 149 of Ref. \citen{bishop2006pattern}. The existing results are primarily developed for supervised regression tasks, but here we re-interpreted them in the context of our unsupervised SSF reconstruction task.

Specifically, denote the set of parameters that can best represent the SSF data as ${\bs \Theta}^*$, i.e., ${\bs \Theta}^*=\min_{{\bs \Theta}}\|\bc{X}-\bc D( \bs \Theta)\|_\F^2$. The reconstruction error can be shown (see Appendix \ref{appendix-b}) to be decomposed as
\begin{align}
    E= E_1+E_2+\epsilon,
    \label{eq:ed}
\end{align}
where $E_1=\|\bc{X}-\bc D( {\bs \Theta}^*)\|_\F^2$ is the representation error; $E_2=\|\bc D(\hat{\bs \Theta})-\bc D( {\bs \Theta}^*)\|_\F^2$ is the identification error; and $\epsilon$ is the cross term defined as $\epsilon=2\langle \bc{X}-\bc D( {\bs \Theta}^*), \bc D( {\bs \Theta}^*)-\bc{D}(\hat{\bs \Theta})\rangle$.

From the definition of the three terms in Eq.~\eqref{eq:ed}, it is easy to show  (see Appendix \ref{appendix-b}) that the cross term $\epsilon$ is equal to zero if $E_1$ or $E_2$ is equal to zero, i.e.,
\begin{align}
    \epsilon=0\Leftarrow \bc{X}=\bc D( {\bs \Theta}^*)~\text{or}~\bc D( {\bs \Theta}^*)=\bc D(\hat{\bs \Theta}).
    \label{eq:cross}
\end{align}


The error decomposition results show that minimizing the reconstruction error $E$ requires simultaneously reducing the representation error $E_1$ and the identification error $E_2$.  Like neural networks, models with high expressive powers usually have low representation errors but high identification errors. Conversely, concise models, like linear regression,  have low identification errors but high representation errors due to their limited parameters and simple operations.

The result in \eqref{eq:cross} suggests two possible ways to minimize the reconstruction error. To null the cross term $\epsilon$, one can choose a complicated model with universal approximation property \cite{hornik1989multilayer} to zero the representation error; or a relatively simple model to induce zero identification error. In the first case, the reconstruction error is up to the identification error, thus necessitating the model complexity optimization. In the latter case, the effort should be paid to enhance the expressive power of the model while not increasing the model's complexity. 

To draw more insights into the reconstruction error, two commonly used reconstruction methods are analyzed in the following subsection.

\subsection{Commonly Used Approaches}
\subsubsection{\label{Sec II-A} Spline Interpolation} 
The biharmonic spline interpolation assumes that the SSF can be well represented by a linear combination of smooth Green functions centered at the samples, i.e.,\cite{sandwell1987biharmonic}
\begin{align}
\bc{X}_{\mathbf{i}}=[\bc D(\mathbf{w})]_{\mathbf{i}}=\sum_{n=1}^Nw_ng(\mathbf{i},\mathbf{i}_n),
\end{align}
where $\mathbf{i}_n = (i_n, j_n, k_n) \in \mathbb{R}^{3}$ is the index of the sampled location; $w_n$ is the weight associated with each sample; $\bs \Theta=\mathbf{w}=[w_1,\cdots,w_N]^\T$; and $g(\cdot,\cdot)$ is the Green function. The representation error $E_1$ is given by
\begin{align}
    E_1=\|\bc{X}-\bc D(\mathbf{w}^*)\|_\F^2,
\end{align}
where the optimal weights $\mathbf{w}^*$ are determined by solving the following problem:
\begin{align}
    \min_{\mathbf{w}}\sum_{n=1}^N\left[\bc{Y}_{\mathbf{i}_n} - \sum_{k=1}^Nw_kg(\mathbf{i}_n,\mathbf{i}_k) \right]^2.
    \label{eq:spline}
\end{align}
Note that the optimal weights can be determined uniquely from Eq.~\eqref{eq:spline} since it is a well-posed problem, which leads to zero identification error, i.e., $E_2=0$. However, the predefined Green function is not flexible and may not reflect the correlations among the sound speeds at different regions, resulting in a significant $E_1$. According to the results in \eqref{eq:ed} and \eqref{eq:cross}, since $E_2=0$ and $\epsilon=0$, the reconstruction error $E = E_1$.

\subsubsection{\label{Sec II-B} Deep Neural Networks} 
In recent years, deep neural networks (DNN) have become highly effective tools in various areas, including computer vision,\cite{he2022masked} natural language processing, \cite{wolf2020transformers} and acoustic signal processing.\cite{adavanne2018sound,hua2023interpretable,ozanich2020feedforward}  
Here we assume that the SSF can be represented by a $L$-layer neural network, i.e.,
\begin{align}
    \bc{X}=\bc D(\bs \Theta)=F_{\bs \theta_L}(F_{ \bs \theta_{L-1}}(\cdots F_{\bs \theta_1}(\bc{G}) )),
\end{align}
where $F_{\bs \theta_l}(\cdot)$ denotes the function of the $l$th layer with parameters (e.g., weights and biases) $\bs \theta_l$; and $\bc{G}$ is the core tensor. The parameter set $\bs \Theta$ includes $\bc{G}$ and  $\{ \bs \theta_l \}_l$. For the functional forms of  $\{F_{\bs \theta_l}(\cdot)\}_l$, commonly used ones include the multi-layer perceptron (MLP)\cite{hastie2009elements} and convolutional neural network (CNN).\cite{o2015introduction} 

For the DNN-based SSF representation model, the identification error is given by
\begin{align}
    E_2=\|F_{\bs \theta^*_L}(F_{ \bs \theta^*_{L-1}}(\cdots F_{\bs \theta^*_1}(\bc{G}^*) ))-F_{\hat{\bs \theta}_L}(F_{  \hat{\bs\theta}_{L-1}}(\cdots F_{\hat{\bs \theta}_1}(\hat{\bc{G}}) ))\|_\F^2
\end{align}
where $\bs{\Theta}^*$ (includes $\bc{G}^*$ and $\{\bs{\theta}^*_l\}_l$) is the solution to the representation problem
\begin{align}
    \min_{\bs{\Theta}}\|\bc{X}-F_{\bs \theta_L}(F_{ \bs \theta_{L-1}}(\cdots F_{\bs \theta_1}(\bc{G}) ))\|_\F^2;
\end{align}
and $\hat{\bs \Theta}$ (includes $\hat{\bc{G}}$ and $\{\hat{\bs{\theta}_l}\}_l$) is the solution to the reconstruction problem 
\begin{align}
    \min_{\bs{\Theta}}\|\bc{Y}-\bc{O}*F_{\bs \theta_L}(F_{ \bs \theta_{L-1}}(\cdots F_{\bs \theta_1}(\bc{G}) ))\|_\F^2.
    \label{eq:nnrec}
\end{align}
Due to the universal approximation property \cite{hornik1989multilayer} of DNN, the representation error $E_1$ can be zero, i.e., $E_1= 0$, if a sufficient number of neurons are available and/or a highly sophisticated neural architecture is utilized. According to the results in \eqref{eq:ed} and \eqref{eq:cross}, since $E_1=0$ and $\epsilon=0$, the reconstruction error $E = E_2$, which, however, will be large since the solution space of problem \eqref{eq:nnrec} is vast (that is, there exists a large number of solutions that attain the same objective value of problem~\eqref{eq:nnrec}).

\section{\label{sec:3} Tensor Neural Network-aided Reconstruction}

The reconstruction error analysis introduced in the last section shows that a highly expressive representation model is necessitated to accurately capture the complex details of sound speeds' spatial distribution, thus minimizing the representation error. But this often leads to increased parameters and complicated mathematic operations, resulting in a more complicated model with increased identification error. Consequently, the key to approaching the optimal reconstruction is to enhance the model's expressive power while maintaining the model's conciseness (in terms of both parameter number and mathematical operations). Towards this goal, in this section, inspired by the recent success of succinct tensor-based ocean SSF representation and powerful deep neural network-based image representation, we {\it take the best from them} and propose a 3D SSF-tailored tensor deep neural network. Then, an effective reconstruction algorithm is derived. 

In the following subsections, we first briefly review the recent tensor-based SSF representation model (Sec.~\ref{Sec III-A}), and then devise our proposed model that integrates the tensor model and deep neural network (Sec.~\ref{Sec III-B}). Next, theoretical insights are drawn by analyzing the proposed model's noise rejection property and particular form (Sec.~\ref{Sec III-C}). Finally, the reconstruction algorithm is derived  (Sec.~\ref{Sec III-D}).  

\subsection{\label{Sec III-A}  Tensor-based SSF Representation} 
This subsection briefly reviews the tensor-based SSF representation framework established in Ref.~\citen{cheng2022tensor}, which relies on the Tucker decomposition model.  For readers unfamiliar with tensors, a brief review in the context of SSF representation was provided in Sec. III of Ref.~\citen{cheng2022tensor}. The definitions of the tensor operations utilized in this paper are elucidated in Appendix \ref{appendix-a}. 

Mathematically, the tensor-based SSF representation model is:
\begin{align}
\bc{X} \approx \bc{S} \times_1 \mathbf{U}^{(1)} \times_2 \mathbf{U}^{(2)} \times_3 \mathbf{U}^{(3)},
\label{eq:tucker}
\end{align}
where the columns of factor matrices $\{\mathbf{U}^{(p)}\}_{p=1}^3$ can be interpreted as the spatial basis functions; and core tensor $\bc{S}$ contains the weighting coefficients. Symbol  $\times_p$ denotes the mode-$p$ product; see the definition in Appendix \ref{appendix-a}. Various constraints can be specified on factor matrices and the core tensor to incorporate additional knowledge, e.g., orthogonal constraints\cite{cheng2022tensor} and sparsity constraints.\cite{chen2022tensor}

Ref.~\citen{cheng2022tensor} has shown the conciseness of the tensor-based representation model, which is mathematically succinct and has a small number of parameters, leading to a small identification error $E_2$. On the other hand, the tensor-based representation model is also effective in characterizing the three-dimensional interactions among sound speeds. Notably, it includes the classical SSF basis functions, such as EOFs and Fourier basis functions, as special cases. \cite{cheng2022tensor} However, since the Tucker decomposition model is based on a multi-linear form, the representation model in Eq.~\eqref{eq:tucker}  is difficult to capture the highly nonlinear variations of sound speeds (caused by small-scale/mesoscale ocean processes), resulting in a non-negligible representation error $E_1$, as demonstrated by the numerical results presented in Sec.~\ref{Sec IV-B}.

\subsection{\label{Sec III-B}  Integrating Tensor and Neural Networks for SSF Representation} 
\begin{figure}
    \center
    \includegraphics[width=0.9\reprintcolumnwidth]{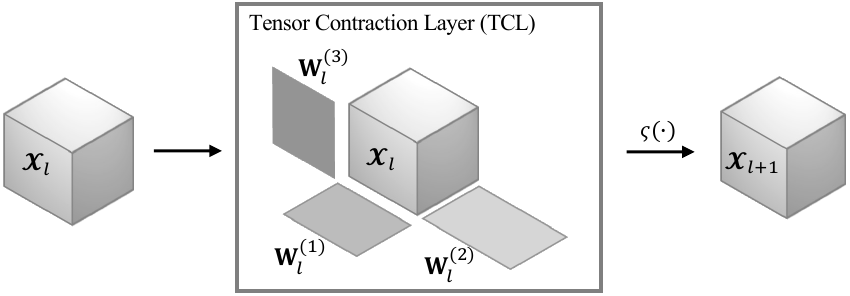}
    \caption{ A graphical illustration of the tensor contraction layer (TCL). $\bc{X}_l$ is the input; $\bc{X}_{l+1}$ is the output; $\{\mathbf{W}^{(i)}_l\}_{i=1}^3$ are the factor matrices; and $\varsigma(\cdot)$ is the activation function.  }
    \label{fig:tcl}
    \hrule
\end{figure}

In this subsection, we propose an integrated model that can take both advantages of DNN and tensor-based representation models, i.e., expressiveness and conciseness, respectively. The proposed model is based on the tensor contraction layer (TCL),\cite{kossaifi2020tensor} as shown in  Fig.~\ref{fig:tcl}. 


Given an input tensor $\bc{X}_{l}\in\mathbb{R}^{R_1^{l}\times R_2^{l}\times R_3^{l}}$, the output of TCL is:
\begin{align}
        \bc{X}_{l+1}= \varsigma(\bc{X}_{l}\times_1 \mathbf{W}^{(1)}_l\times_2\mathbf{W}^{(2)}_l\times_3\mathbf{W}^{(3)}_l),
\end{align}
where  the output $\bc{X}_{l+1}\in\mathbb{R}^{R_1^{l+1}\times R_2^{l+1}\times R_3^{l+1}}$; each matrix 
$\mathbf{W}^{(i)}_l \in \mathbb{R}^{R_i^{l+1}\times R_i^{l}} $; and  $\varsigma(\cdot)$ is the activation function. Note that the TCL is based on the Tucker decomposition, enabling it to effectively capture the multidimensional interactions of sound speeds even with a small number of parameters.

\begin{figure}[t]
    \center
    \includegraphics[width=1.05\reprintcolumnwidth]{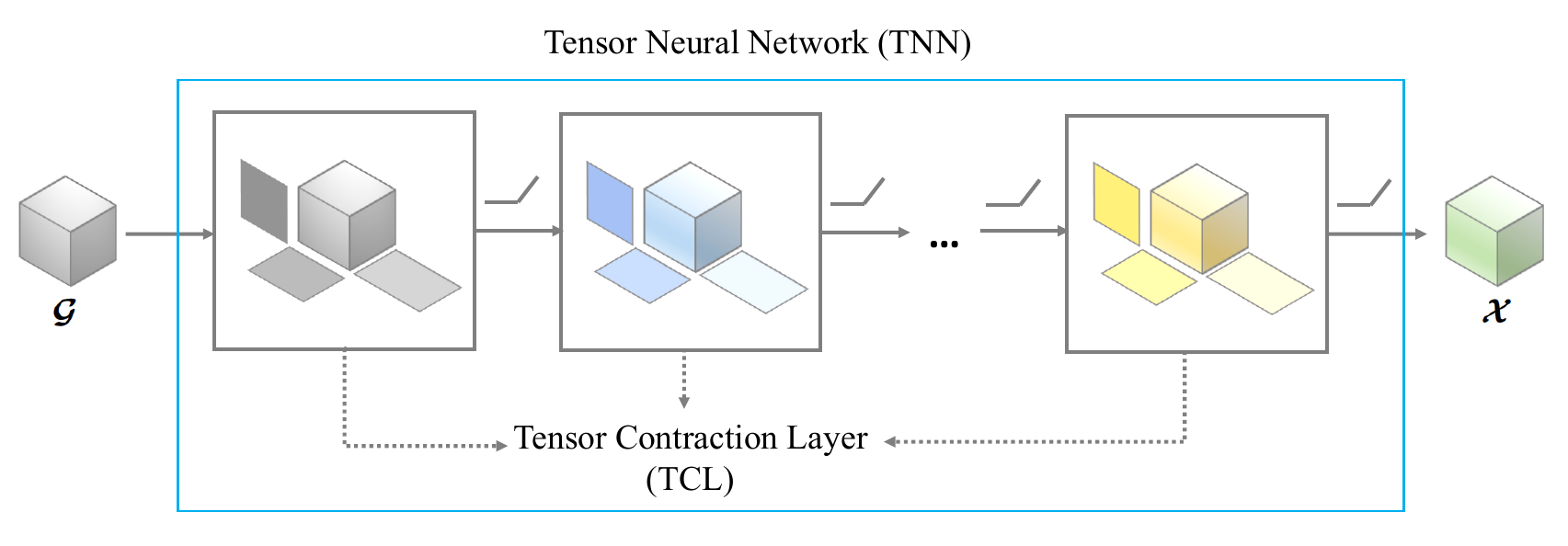}
    \caption{ Illustration of the tensor neural network (TNN), which is composed of multiple TCLs. $\bc{G}$ is the core tensor and $\bc{X}$ is the output tensor.}
    \label{fig:TNN}
    \hrule
\end{figure}

To further enhance the expressive power, we propose a tensor neural network (TNN) composed of multiple TCLs; see Fig.~\ref{fig:TNN}. 
Mathematically, a $L$-layer TNN can be defined as 
\begin{align}
\label{eq:TNN}
\begin{aligned}
\bc{X} = \varsigma\bigg(&\cdots\varsigma\left( \bc{G}\times_1 \mathbf{W}^{(1)}_1 \times_2 \mathbf{W}^{(2)}_1 \times_3 \mathbf{W}^{(3)}_1  \right) \cdots \\
&\times_1 \mathbf{W}^{(1)}_L \times_2 \mathbf{W}^{(2)}_L \times_3 \mathbf{W}^{(3)}_L \bigg),
\end{aligned}
\end{align}
where $\bc{G}\in \mathbb{R}^{R_1 \times R_2 \times R_3}$ is the core tensor, $\{\{\mathbf{W}^{(i)}_l\}_{i=1}^3\}_{l=1}^L$ are the factor matrices. The choice of activation function depends on the data. The commonly used ones are rectified linear unit (ReLU), sigmoid, tanh, etc. \cite{karlik2011performance}

\subsection{\label{Sec III-C}  Theoretical Insights} 
Before deriving the reconstruction algorithm, theoretical analyses were performed to reveal the insights of the proposed TNN-based representation model.

First, we show that the particular architecture of the proposed model enables it to reject noise effectively.  
Here the Gaussian noise is considered due to the least-squares loss function used in Eq.~\eqref{eq1}, which is a common choice in physical field reconstruction research \cite{shrestha2022deep, caviedes2021gaussian}.
For simplicity,  consider a {\it one-layer} TNN equipped with a ReLU activation function. The TNN output can be written as:
\begin{align}
    \bc{X}=\text{ReLU}(\bc{G}\times_1 \mathbf{W}^{(1)}_1 \times_2 \mathbf{W}^{(2)}_1 \times_3 \mathbf{W}^{(3)}_1).
\end{align}
Then we have the following proposition.

 \begin{tcolorbox}
\textbf{Proposition 1}: 
	Consider a {\it one-layer} TNN with parameters
	$\mathbf{\Theta}=\{\bc{G},\mathbf{W}^{(1)}_1,\mathbf{W}^{(2)}_1,\mathbf{W}^{(3)}_1\}$ and a noise tensor $\bc{E}$ with each element following a zero-mean i.i.d Gaussian distribution, i.e., $\bc{E}_{i,j,k}\sim\mathcal{N}(0,\sigma^2)$. Then with probability at least $1-e^{-R}-e^{-\frac{T}{16}}$,
\begin{align}
    \min_{\mathbf{\Theta}}
    \|\bc{X}-\bc{E}\|_\F^2\geq \|\bc{E}\|_\F^2(1-\frac{10R}{T}),
    \label{eq:nfit}
\end{align}
where $R=R_1R_2R_3$ and $T=IJK$. \\
{\it Proof}: See Appendix \ref{appendix-b}. 
\end{tcolorbox}

The term  $\|\bc{X}-\bc{E}\|_\F^2$  in Eq.~\eqref{eq:nfit} serves as a metric for assessing the model's capacity to fit the noise. A higher value of this term indicates that the model is less prone to fit the noise. \textbf{Proposition 1} provides a lower bound for  $\|\bc{X}-\bc{E}\|_\F^2$, which is a product of the noise power $\|\bc{E}\|_\F^2$   and a constant  $(1-\frac{10R}{T})$. Since $R$ is typically much smaller than $T$ to promote model conciseness  (i.e., $R  \ll T$), the constant $(1-\frac{10R}{T})$ is approximately equal to 1. As a consequence, the lower bound is close to the noise power, which is significantly greater than zero. Therefore, \textbf{Proposition 1} points out that the one-layer TNN rejects fitting the noise with a high probability. 
Note that different from classical low-rank tensor decomposition works (Refs.~\citen{kolda2009tensor,sidiropoulos2017tensor,de2000multilinear} and references therein) that are primarily based on multi-linear forms, \textbf{Proposition 1} theoretically quantifies the noise-fitting ability of the proposed tensor model with the ReLU nonlinear activation function,  which is novel and has a standalone value beyond SSF reconstruction tasks.

The key concept underlying \textbf{Proposition 1} is the intuition that a non-linear tensor decomposition model with intrinsically low-rank property, such as the one defined in Eq.~\eqref{eq:TNN}, can effectively reject noise. 
While this proposition is  specifically formulated for the {\it one-layer} TNN, the intuition behind noise rejection extends to TNN models with multiple layers. However, extending the results in {\bf Proposition 1} to multiple-layer TNNs poses a daunting challenge. As a result, we experimentally validate its noise rejection property instead.

\begin{figure}
    \center
    \includegraphics[width=1\reprintcolumnwidth]{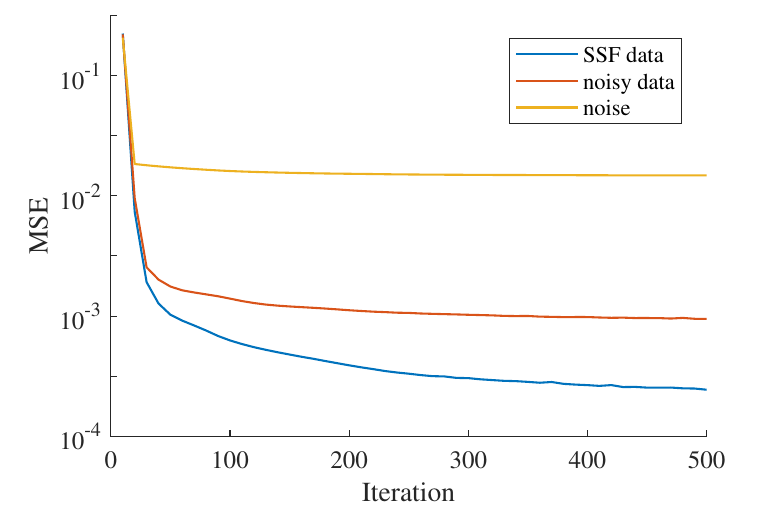}
    \caption{ The mean squared error (MSE) curves were obtained by fitting the proposed model to white Gaussian noise with standard deviation $\sigma = 0.5$, SSF data, and SSF data with added noise. The three-layer TNN model used in this experiment has the same architecture as the one used in Sec.~\ref{sec:4} (see Fig.~\ref{fig:tnn}). The results demonstrate that the proposed model is inclined to fit the SSF data while effectively rejecting noise fitting.}
    \label{fig:losses}
    \hrule
\end{figure}

The multiple-layer TNN model used in this experiment has the same architecture as the one used in Sec.~\ref{sec:4} (see Fig.~\ref{fig:tnn}), which has three layers with dimensions being (5, 5, 5), (10, 10, 10) and (20, 20, 20) respectively.
In our experiment, we fit the proposed model to three distinct datasets: white Gaussian noise with standard deviation $\sigma = 0.5$, SSF data, and SSF data with added noise. In Fig.~\ref{fig:losses}, we present the fitting error, quantified by the mean squared error (MSE), with respect to the iteration number of the parameter learning process. Fig.~\ref{fig:losses} demonstrates that the proposed model effectively fits the SSF data, as evidenced by the smallest MSE values across all iteration numbers. Conversely, the MSE values of the noise fitting are the highest, substantiating the model's noise rejection property. Furthermore, when the SSF data is contaminated by noise, the proposed model tends to fit the SSF data while fitting only a minor portion of the noise, resulting in the MSE curve (red curve) being much closer to that of fitting the noise-free SSF data (blue curve).

Subsequently, we establish a connection between the proposed TNN and the tensor Tucker model defined in Eq.~\eqref{eq:tucker}. 
Specifically, if the activation function $\varsigma(\cdot)$ is chosen to be the identity or linear function, following the definition of tensor Tucker decomposition,\cite{kolda2009tensor,de2000multilinear}  the $L$-layer TNN (defined in Eq.~\eqref{eq:TNN}) can be easily shown (see Appendix \ref{appendix-b}) to be  equivalent to the Tucker decomposition model (defined in Eq.~\eqref{eq:tucker}), that is, 
\begin{align}
\label{eq:tnntucker}
\begin{split}
\bc{X} &= \varsigma\left(\cdots\varsigma\left( \bc{G}\times_1 \mathbf{W}^{(1)}_1 \times_2 \mathbf{W}^{(2)}_1 \times_3 \mathbf{W}^{(3)}_1  \right) \cdots \right. \nonumber\\
&\quad \left. \times_1 \mathbf{W}^{(1)}_L \times_2 \mathbf{W}^{(2)}_L \times_3 \mathbf{W}^{(3)}_L\right),
\end{split} \\
\nonumber \\
\begin{split}
&= \bar{\bc{G}}\times_1 \bar{\mathbf{W}}^{(1)} \times_2 \bar{\mathbf{W}}^{(2)} \times_3 \bar{\mathbf{W}}^{(3)},
\end{split}
\end{align}
where $\varsigma(x)=ax$ with $a\in\mathbb{R}$ being a constant, $\bar{\bc{G}}=a^L\bc{G}$ and $\bar{\mathbf{W}}^{(i)}=\mathbf{W}^{(i)}_L\mathbf{W}^{(i)}_{L-1}\cdots\mathbf{W}^{(i)}_1, i=1,2,3$.

This result reveals that the tensor Tucker decomposition model, which is the cornerstone of the recently developed tensor-based SSF basis function learning framework that includes classical ones (such as EOFs and Fourier basis functions) as special cases, is {\it a particular instance} of the proposed TNN model. In other words, {\it the proposed TNN model is the most general ocean SSF representation model to date}. The introduction of nonlinear activation functions significantly improves the tensor model's representation capability without compromising its conciseness, allowing simultaneous decreases in both the representation error $E_1$ and identification error $E_2$. Specifically, the nonlinearity and hierarchical construction enable the effective capture of sound speed variations, leading to a small $E_1$. Meanwhile, the Tucker decomposition-based backbone, or TCL, preserves the tensor model's conciseness, leading to a small $E_2$.

After establishing a strong link between the proposed tensor neural network and the tensor Tucker decomposition model, one may contemplate extending the recoverability analysis \cite{gandy2011tensor} using Tucker models under either systematic sampling \cite{prevost2020hyperspectral,sorensen2019fiber} or random sampling \cite{liu2012tensor}. However, due to the involvement of nonlinear activation functions, the analysis becomes a formidable challenge. Instead, in Sec.~\ref{sec:4}, we present a numerical analysis of the recoverability of the proposed model, while preserving its theoretical analysis as an intriguing avenue for future research.

\subsection{\label{Sec III-D}  TNN-aided 3D SSF Reconstruction Algorithm} 
Based on the proposed TNN model, the 3D SSF reconstruction problem can be formulated as
\begin{align}
\begin{split}
\label{eq:tnnrec}
\min_{\bs \Theta} ~~~&\underbrace{\|\bc{Y}-\bc{O}*\bc{X}\|_\F^2+\lambda R(\bc{X})}_{\triangleq f(\bs{\Theta})},
\end{split} \\[2ex]
\begin{split}
\text{s.t.}~~~&\bc{X}=\varsigma\left(\cdots\varsigma\left( \bc{G}\times_1 \mathbf{W}^{(1)}_1 \times_2 \mathbf{W}^{(2)}_1 \times_3 \mathbf{W}^{(3)}_1  \right) \cdots \right. \\
&\quad \left. \times_1 \mathbf{W}^{(1)}_L \times_2 \mathbf{W}^{(2)}_L \times_3 \mathbf{W}^{(3)}_L \right),
\end{split}
\end{align}
where $R(\bc{X})$ is a regularization term that allows the incorporation of side information for further performance enhancement, and $\lambda$ is a hyper-parameter that balances the importance of the data fitting and regularization terms.  Various types of structural information can be incorporated by choosing different $R(\bc{X})$. If  $\lambda = 0$, $R(\bc{X})$ is no longer effective. In the experimental results (Sec.~\ref{sec:IV-C}), we demonstrate using the total variation (TV) regularizer to capture spatial correlations among sound speeds.

The model parameters $\bs\Theta$ can be updated using the calculated gradients denoted as $\nabla_{\bs\Theta} f(\bs\Theta)$ via the gradient descent method. \cite{nesterov2003introductory} This algorithm is widely used in deep learning and is straightforward. The gradients of the loss function with respect to the model parameters $\bs\Theta$ can be obtained by utilizing the automatic differentiation technique in TensorFlow.\cite{abadi2016tensorflow} Appendix \ref{appendix-c} provides some key steps of the derivation of the gradients. The update rule for the model parameters is given as follows:
\begin{align}
    \bs{\Theta}_{k+1}\leftarrow \bs{\Theta}_{k} - \eta_k\nabla_{\bs{\Theta}} f(\bs{\Theta}_k),
\end{align}
where $\bs{\Theta}_k$ and $\bs{\Theta}_{k+1}$ are the model parameters at iterations $k$ and $k+1$, respectively, and $\eta_k$ is the learning rate. It's worth noting that the gradient descent method is a first-order method that can scale well to tackle big data.\cite{fan2014challenges} Advanced acceleration schemes such as AdaGrad, RMSProp, and Adam can be used to further speed up the convergence.\cite{ruder2016overview} We summarize the proposed algorithm in \textbf{Algorithm 1}.
Note that the tensor $\bc{G}$ is a model parameter that is optimized during the reconstruction, while not the input of the reconstruction algorithm.

\begin{table}[]
\begin{tcolorbox}
{\textbf{\color{black} Algorithm 1: TNN-aided Ocean SSF Reconstruction Algorithm}}
\\
\begin{ruledtabular}
\begin{tabular}{c} 
\leftline{\textbf {Input:} $\bc{Y}$, $\bc{O}$, threshold $\alpha$, maximum iteration }\\
\leftline{~~~~~~~~~~~~number $K$.}\\
\leftline{\textbf{Initialize:} $k \leftarrow 0$, $\eta_0$, $\bs{\Theta}_0$.} \\
\leftline{\textbf{While} $k<K$ or $|e|>\alpha$ \textbf{do}} \\
\leftline{ ~~~~Caculate the gradients $\nabla_{\bs{\Theta}} f(\bs{\Theta}_k)$,}  \\ 
\leftline{ ~~~~Update $\bs{\Theta}$ via $\bs{\Theta}_{k+1}\leftarrow\bs{\Theta}_{k} -\eta_k\nabla_{\bs{\Theta}} f(\bs{\Theta}_{k})$,}  \\ 
\leftline{ ~~~~$e \leftarrow f(\bs{\Theta}_{k+1})-f(\bs{\Theta}_{k}$)}  \\
\leftline{ ~~~~$k\leftarrow k+1$,}  \\
\leftline{\textbf{end while}} \\
\leftline{\textbf{Return} $\hat{\bc{X}}=\bc{D}(\bs{\Theta}_k)$.
}
\end{tabular}
\end{ruledtabular}
\end{tcolorbox}
\label{algorithm2}
\end{table}

\section{\label{sec:4} Numerical Results and Discussions}
In this section,  numerical results using real-life ocean 3D SSF data are presented to showcase the encouraging reconstruction performance of the proposed tensor neural network-aided reconstruction method (labeled as TNN). 

\subsection{\label{Sec IV-A} Experimental Settings}

\begin{figure*}[t]
    \center
    \includegraphics[width=1.4\reprintcolumnwidth]{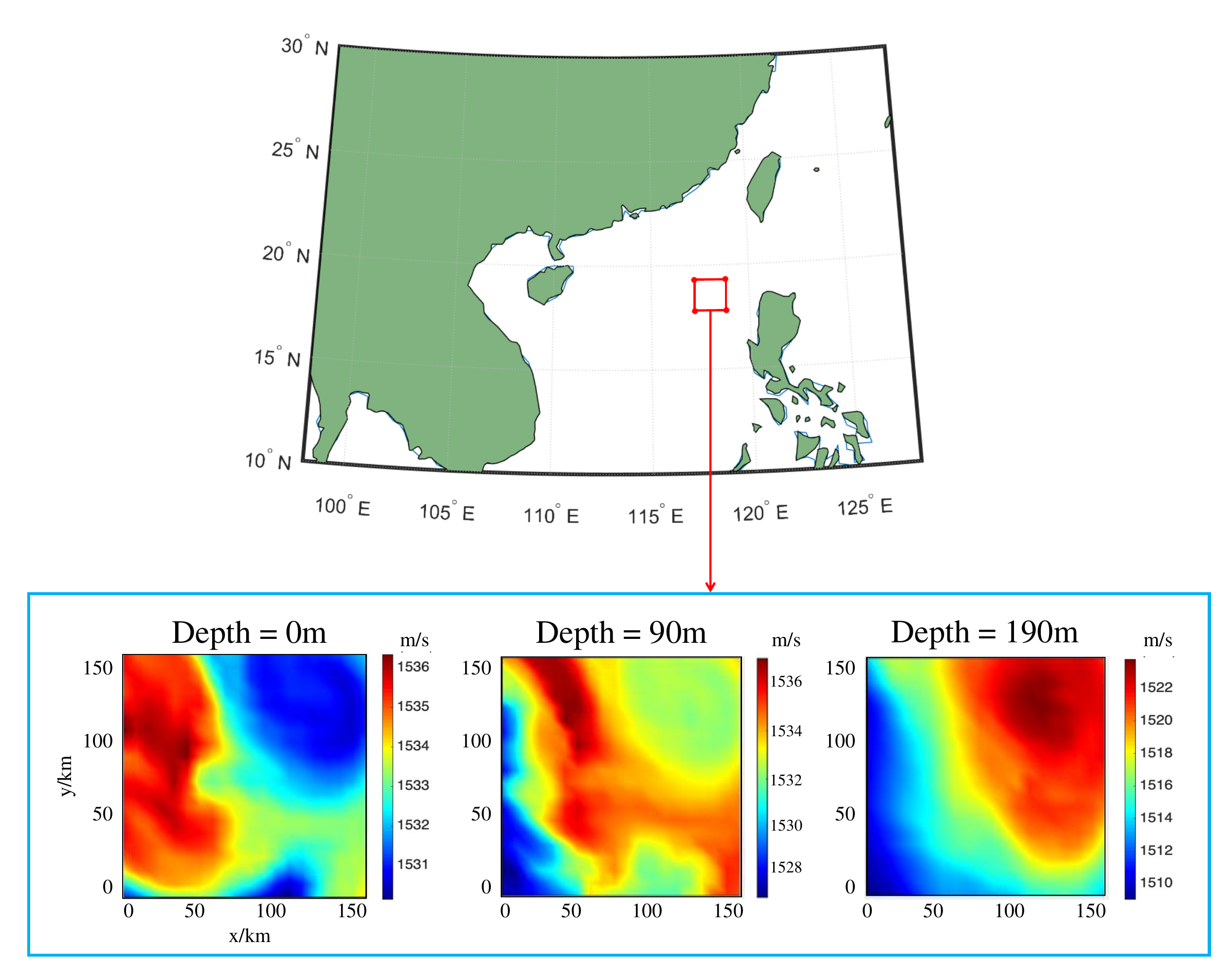}
    \caption{ Illustration of the 3D SSF data.}
    \label{fig:map}
    \hrule
\end{figure*}

\textbf{3D SSF Data}: The 3D South China Sea (SCS) data $\bc{X}\in\mathbb{R}^{20\times 20\times 20}$ on December 21, 2012, is considered in this paper and illustrated in Fig.~\ref{fig:map}. The data were derived from 3D conductivity, temperature, and depth (CTD) measurements using the hybrid coordinate ocean model (HYCOM).  The spatial coverage of the dataset is 152 km $\times$ 152 km $\times$ 190 m, with a horizontal resolution of 8 km and a vertical resolution of 10 m. 

\begin{figure}
    \center
    \includegraphics[width=1\reprintcolumnwidth]{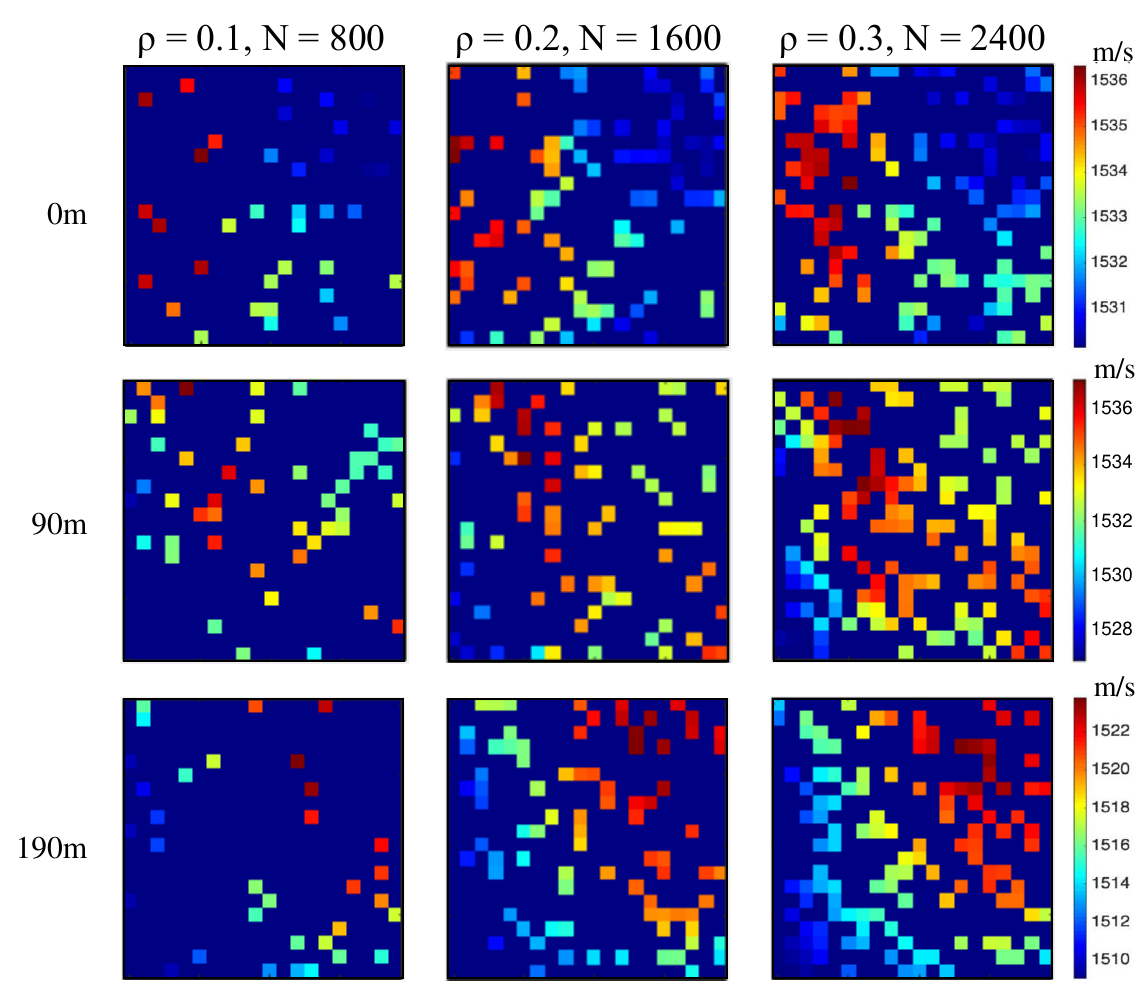}
    \caption{ Measurements of SSF under different sampling ratios in one single Monte-Carlo trial.}
    \label{fig:samples}
    \hrule
\end{figure}

{\textbf{Sound Speed Measurements}}: Following the configuration of the recent ocean observing systems (e.g., the Argo program \cite{nystuen2011interpreted} and Ocean Internet of Things\cite{qiu2019underwater}), the sound speed measurements are randomly collected over the 3D spatial region. 
The sampling ratio is defined as $\rho=\frac{\sum_{i,j,k}\bc{O}_{i,j,k}}{IJK}$. 
Examples of the sampled SSF under different sampling ratios in one single Monte-Carlo trial are shown in Fig.~\ref{fig:samples}. In addition, the sampled data are assumed to be corrupted by i.i.d. Gaussian noise with zero mean and standard deviation $\sigma$. 

\textbf{Performance measure}: The reconstruction performances of different methods are assessed by the root mean squared error (RMSE), given by
\begin{align}
\text{RMSE}=\sqrt{\frac{1}{T}\|\hat{\bc{X}}-\bc{X}\|_\F^2},
\end{align}
where $\bc{X}$ is the ground-truth; $\hat{\bc{{X}}}$ is the reconstructed SSF; and $T=I J K$ denotes the total number of SSF data entries. The RMSEs were averaged over three Monte-Carlo trials with different sampling patterns. All the experiments are conducted in a computer with a 2.2GHz 6-Core Intel i7 CPU.

\textbf{Training Optimizer}: The gradient computation was carried out using the automatic differentiation mechanism of TensorFlow. To optimize the proposed TNN model, we use the popular Adam optimizer with an initial learning rate of 0.005 and fix the number of iterations at 15,000.

\subsection{\label{Sec IV-B} Validating Reconstruction Error Analysis}
To validate the reconstruction error analysis results presented in Sec. \ref{sec:2}, numerical results are first presented in this subsection before comparing the proposed algorithm with SOTA methods. Specifically, we compare the reconstruction performance of the proposed tensor neural network-aided algorithm with the methods that use only deep learning models or tensor models. The results demonstrate the importance of striking the right balance between model conciseness and expressiveness for achieving superior reconstruction performance. In this subsection, no extra prior information is included in the TNN for a fair comparison, i.e., the regularization parameter $\lambda$ is set to zero. And unless otherwise specified, $\sigma$ is set to 0.1, indicating that the RMSEs of sound speed measurements/estimates mostly lie in the range of 0.1-0.3 m/s, a moderate range according to our real-life sea trials. The reconstruction performances under a wider range of noise powers are reported in the next subsection.

\subsubsection{Conciseness Matters}

\begin{figure*}[!t]
    \baselineskip=12pt
    \figline{
    \fig{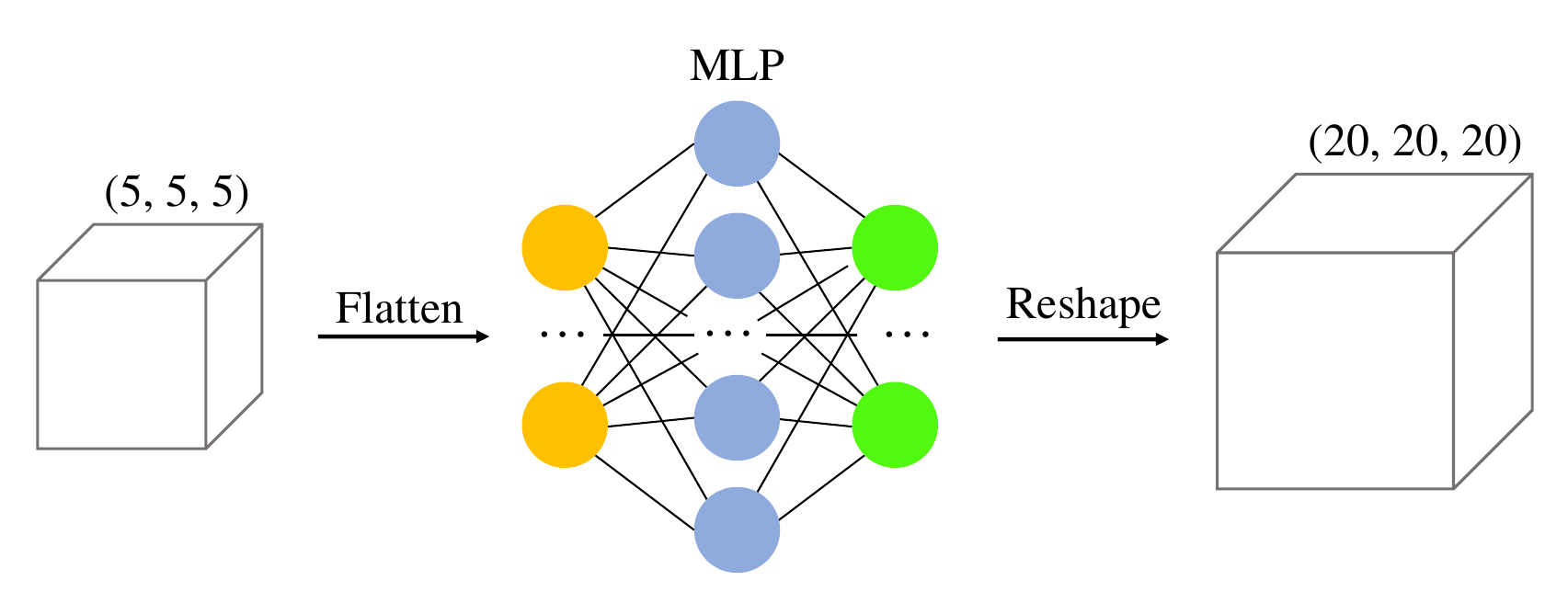}{0.9 \reprintcolumnwidth}{(a)} \label{fig:mlp}
    \fig{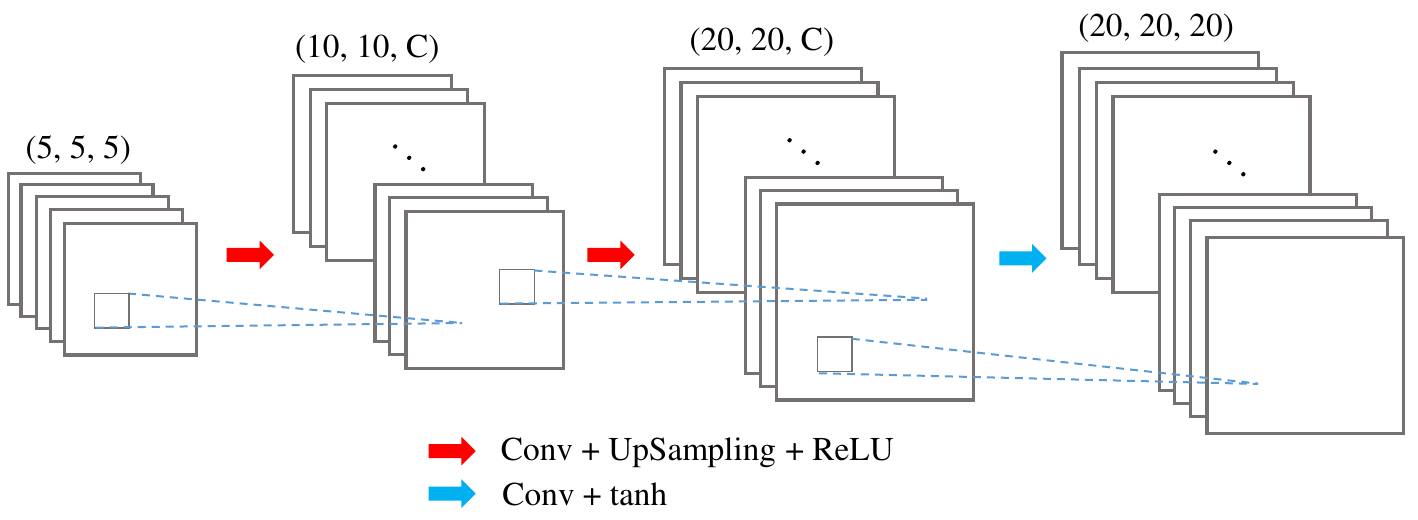}{1 \reprintcolumnwidth}{(b)}  \label{fig:cnn}
    }
    \centering
    \figline{\fig{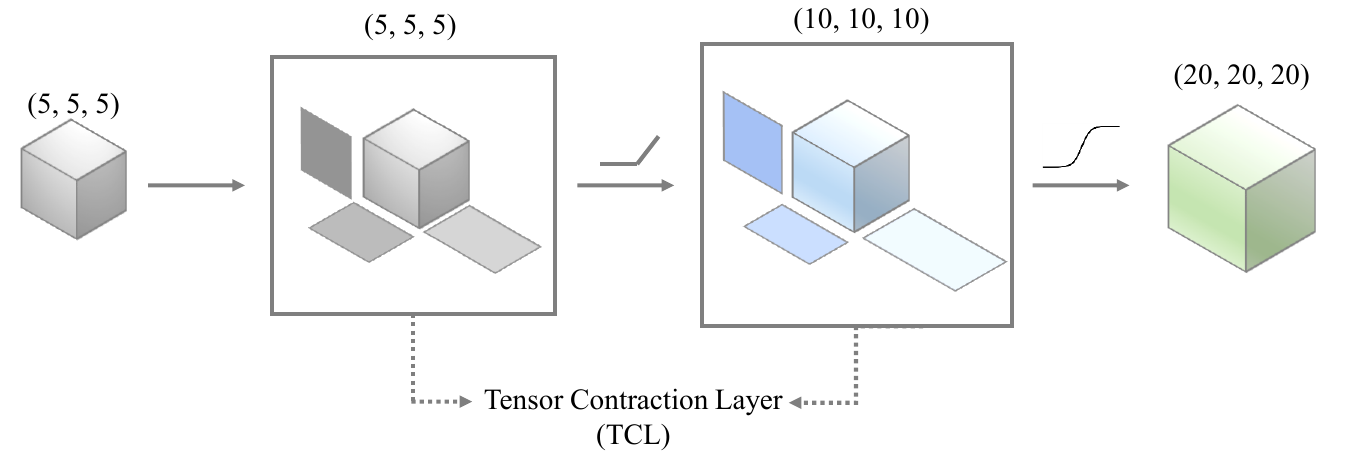}{1.2 \reprintcolumnwidth}{(c)}  \label{fig:tnn}}
    \caption{\label{fig:nn} Th network architecture of (a) MLP, (b) CNN, (c) TNN. The number of channels C is set to 10.    The MLP is composed of three layers with dimensions of 125, 1000, and 8000, respectively. The convolutional layers have a kernel size of (2,2), and the Upsampling layers use the nearest-neighbor interpolation method. For all neural networks, the ReLU activation function is chosen for all layers except for the final layer, which uses the tanh activation function.
    }
    \hrule
\end{figure*}

\begin{figure}
    \center
    \includegraphics[width=1\reprintcolumnwidth]{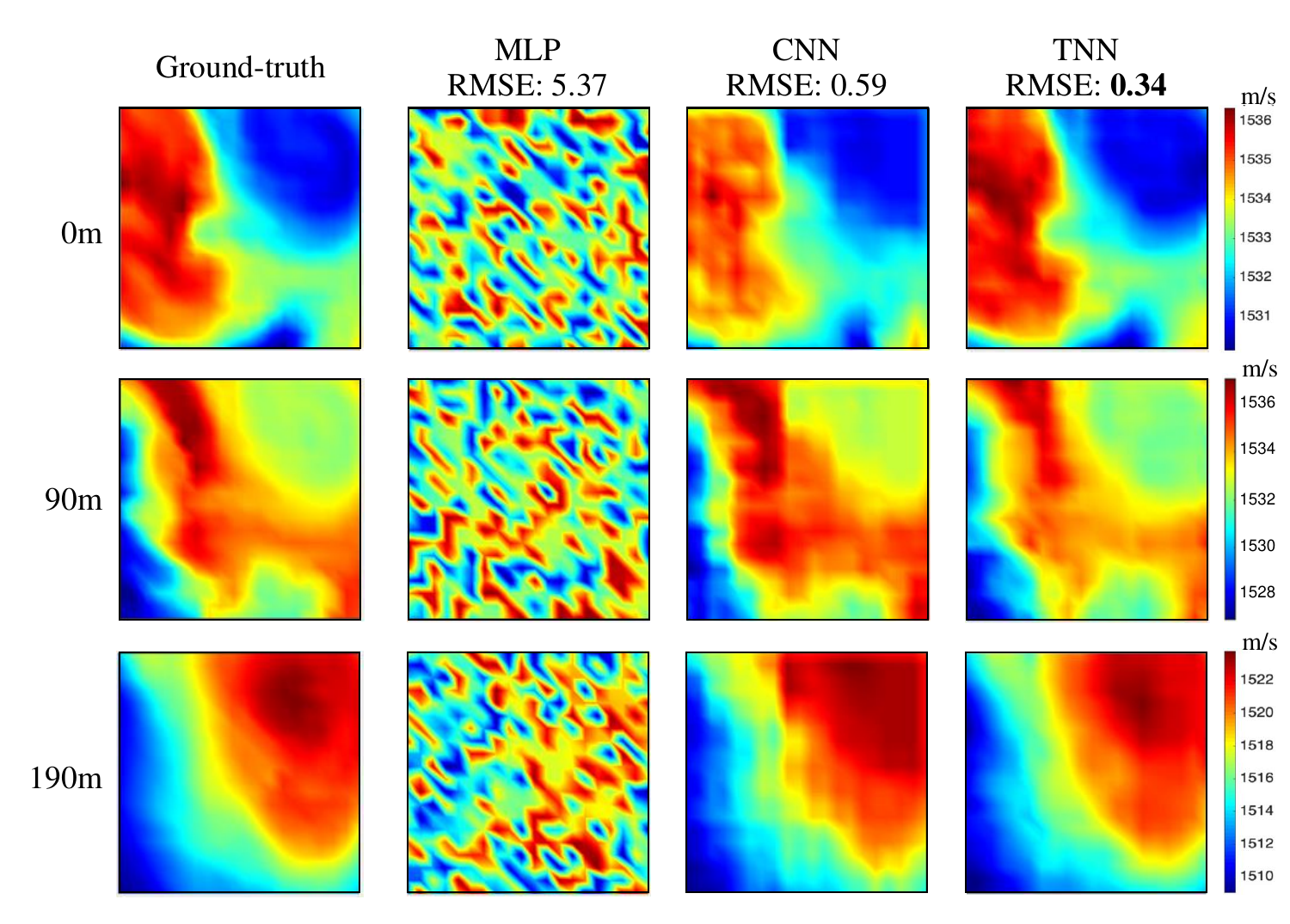}
    \caption{ Visual effects of reconstructed SSF of different neural network-based methods at depths 0 m, 90 m, and 190 m under a sampling ratio of 0.3 in one single Monte-Carlo trial. The RMSEs are shown at the top.}
    \label{fig:nnresluts}
    \hrule
\end{figure}

First, the importance of conciseness in reconstruction is demonstrated by comparing the reconstruction performances of different neural network-based methods. The baseline networks used for this comparison were a three-layer multi-layer perceptron (MLP) and a three-layer convolutional neural network (CNN). In addition, a three-layer tensor neural network (TNN) was used for the proposed method. The detailed network architectures are illustrated in Fig.~\ref{fig:nn}. The reconstruction performances of these network-based methods under different sampling ratios are presented in Table~\ref{tab:nn}. For visualization, Fig.~\ref{fig:nnresluts} presents the recovered SSF of different methods under a sampling ratio of 0.3 in one Monte-Carlo trial.


The results presented in Table~\ref{tab:nn} and  Fig.~\ref{fig:nnresluts} indicate that the SSF reconstruction performance of the MLP is poor. This is due to the large number of parameters in the MLP (around 8 million), which make it difficult to learn the optimal model from limited and noisy measurements (e.g., 2400), resulting in a significant identification error $E_2$. In contrast, the CNN, which has fewer parameters (1565), can leverage the spatial smoothness of the SSF and exhibit better reconstruction performance. However, the CNN does not exploit the 3D spatial correlations among sound speeds as well as the TNN. Specifically, the proposed TNN model, utilizing tensor computation, can effectively capture the 3D spatial correlations with fewer parameters (875), resulting in significantly better reconstruction performance.

\begin{table}[!t]
    \caption{ \label{tab:nn} RMSEs of different  neural network-aided methods under different sampling ratio $\rho$. The number of measurements $N$ is presented in the second column. And the number of parameters for different models is in parentheses. }
    \begin{ruledtabular}
        \begin{tabular}{|c|c|c|c|c|}
        $\rho$ & $N$ & \textbf{MLP}($\approx$ 8M) & \textbf{CNN}(1565) & \textbf{TNN}(875) \\ \hline
        0.2 & 1600 & 5.67 & 0.63 & \textbf{0.50} \\ \hline
        0.3 & 2400 & 5.37 & 0.59 & \textbf{0.34}  \\ \hline
        0.4 & 3200 & 4.94 & 0.56 & \textbf{0.28}
        \end{tabular}
    \end{ruledtabular}
\end{table}

\begin{figure}[t]
	\centering
	\includegraphics[width=0.9\linewidth]{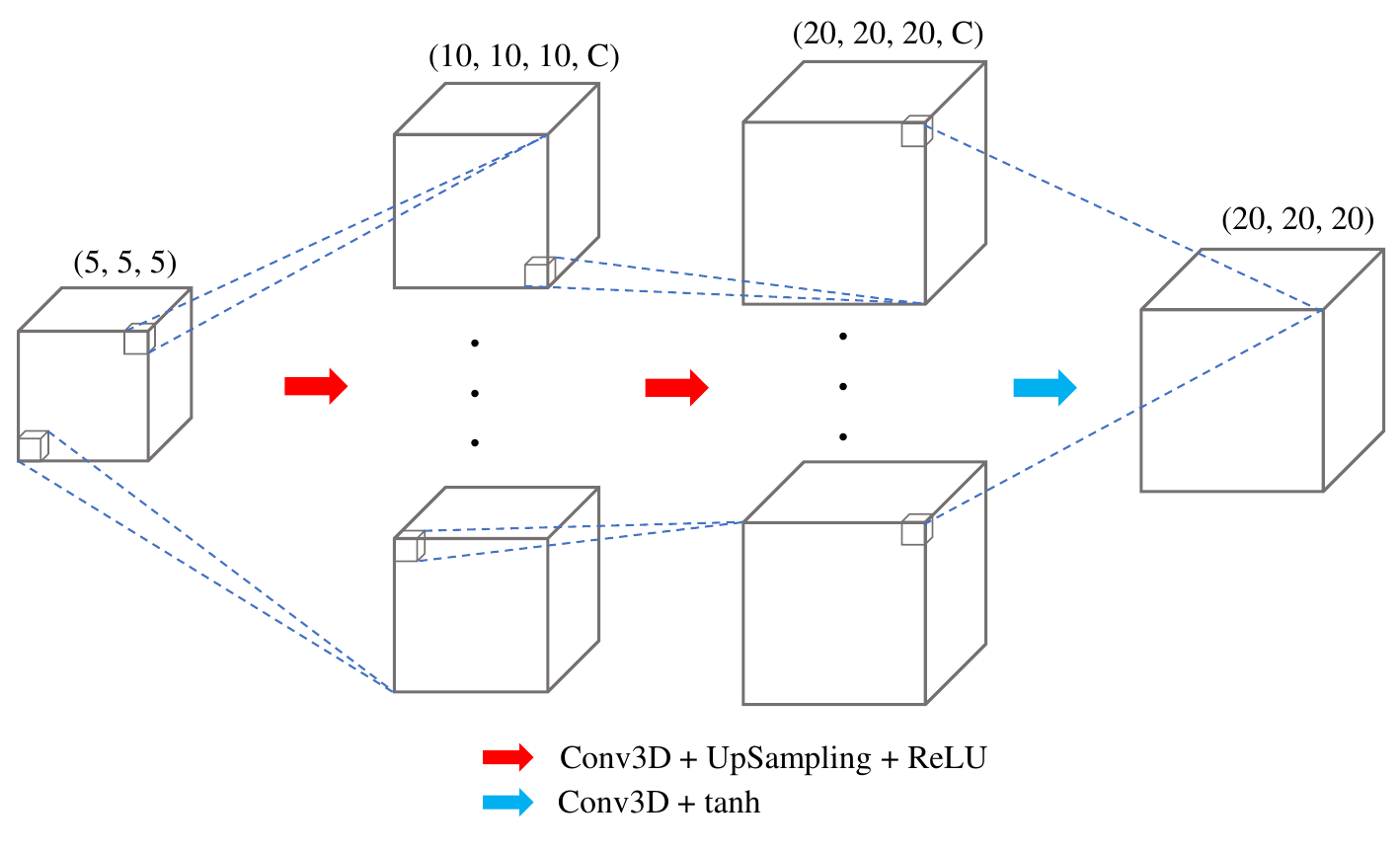}
	\caption{The architecture of the 3D CNN model, where $C$ is the number of kernels in the hidden layer.}
	\label{fig:3DCNN}
\end{figure}

 Additionally, we compare the reconstruction performance of TNN and CNN with 3D convolutions. 
 Fig.~\ref{fig:3DCNN} shows the architecture of the three-layer CNN model with 3D convolutions, while the three-layer TNN model has the same neural network architectures as the one in Fig.~\ref{fig:tnn}. 
Here we denote the CNN models with $C=10$ and $C=20$ as CNN1 and CNN2 respectively, where $C$ is the number of kernels in the hidden layer.
Table~\ref{tab:3dcnn} presents the averaged reconstruction RMSEs of different models under various sampling ratios. Our findings are as follows:

1) TNN achieves the best performance in all scenarios. While CNN models are capable of exploiting spatial correlations, they do not fully leverage the intrinsic low-rankness of the SSF. On the other hand, the TNN model, which nonlinearly concatenates several low-rank tensor models, better exploits the global coherence of the SSF. Notably, even with the minimum number of model parameters, the TNN model achieves the best reconstruction performance among the three models when the SSF is fully observed ($\rho = 1$).
This result indicates that the TNN model has a lower representation error because the identification error is negligible when $\rho = 1$, demonstrating the strong expressiveness of the TNN model for SSF representation.

2) CNN2 outperforms CNN1 in all scenarios. CNN2 has more parameters and higher expressive power, resulting in a lower representation error. Surprisingly, CNN2 also yields better reconstruction results under sparse sampling conditions (e.g., $\rho=0.3$). This could be attributed to the inductive bias of the network architecture, allowing it to produce reasonable outcomes even for highly ill-posed inverse problems, as evidenced by recent works in the deep learning literature (see Ref.~\citen{heckel2020compressive}).

\begin{table}[t]
\centering
\caption{The averaged RMSEs of TNN and 3D CNN models under different sampling ratios. The number of model parameters is in parentheses.}
\label{tab:3dcnn}
\begin{ruledtabular}
\begin{tabular}{|c|c|c|c|}
$\rho$ & TNN(875) & CNN1(1106) & CNN2(3686) \\
\hline
0.3 & \textbf{0.35} & 0.57 & 0.46 \\
\hline
0.5 & \textbf{0.23} & 0.51 & 0.40 \\
\hline
1 & \textbf{0.21} & 0.48 & 0.37
\end{tabular}
\end{ruledtabular}
\end{table}

\subsubsection{Expressiveness Matters}

\begin{table}[t]
    \caption{ \label{tab:tucker} RMSEs of different tensor-based methods under different sampling ratios. The number of measurements $N$ is presented in the second column. And the number of parameters for different models is in parentheses.}
    \begin{ruledtabular}
        \begin{tabular}{|c|c|c|c|c|}
        $\rho$ & $N$ & \textbf{Tucker1}(875) & \textbf{Tucker2}(908) & \textbf{TNN}(875) \\ \hline
        0.2 & 1600 &\textbf{0.47} & 1.20 & 0.50 \\ \hline
        0.3 & 2400 & 0.43 & 0.70 & \textbf{0.34}  \\ \hline
        0.4 & 3200 &0.42 & 0.34 & \textbf{0.28}
        \end{tabular}
    \end{ruledtabular}
\end{table}

Then, experiments were conducted to demonstrate the effectiveness of enhancing the model's representation capability (expressiveness) in improving reconstruction performance. We compare the proposed model with two tucker models with different architectures. The first model, named Tucker1, has the same structure as the TNN model but without non-linearity (i.e., using linear activation functions). The second model, named Tucker2, is just a tensor Tucker decomposition model (defined in Eq.~\eqref{eq:tucker}) with the dimension of the core tensor setting to $(7,8,8)$. 

\begin{figure}[t]
    \center
    \includegraphics[width=1\reprintcolumnwidth]{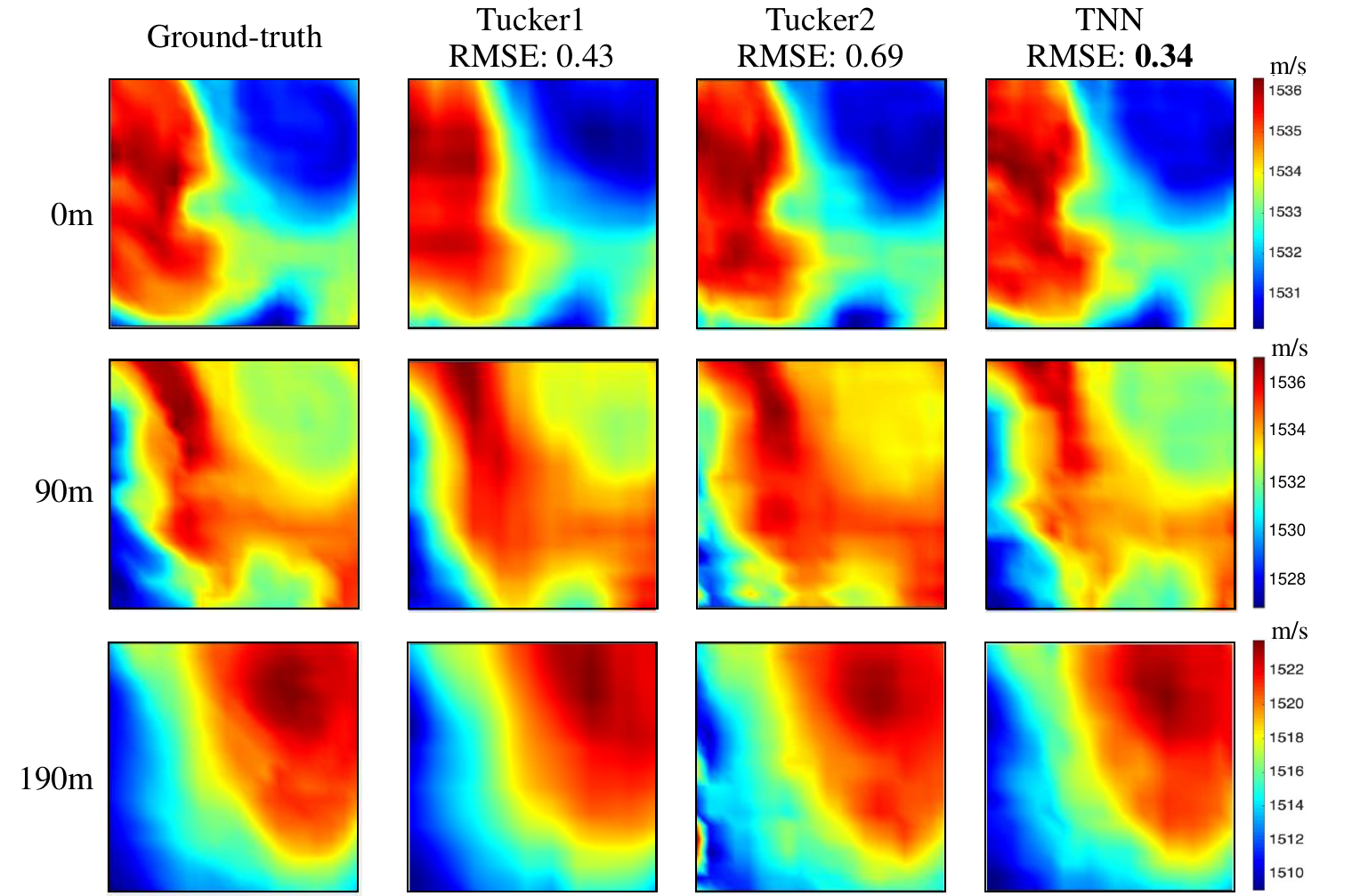}
    \caption{ Visual effects of reconstructed SSF of different tensor-based models at depths 0 m, 90 m, and 190 m under a sampling ratio of 0.3 in one single Monte-Carlo trial. The RMSEs are shown at the top.}
    \label{fig:tuckerresluts}
    \hrule
\end{figure}

The number of parameters in Tucker2 is $20\times(7+8+8)+7\times8\times8=908$, while TNN and Tucker1 both have $(20\times10+10\times5)\times3+5\times5\times5=875$ parameters. Therefore, the three models have a similar number of parameters.   Table~\ref{tab:tucker} presents the average RMSEs of the three models under different sampling ratios. Visualization of the reconstructed SSFs in one Monte-Carlo trial under $\rho=0.3$ is shown in Fig.~\ref{fig:tuckerresluts}.

Based on the experimental results shown in Table~\ref{tab:tucker} and Fig.~\ref{fig:tuckerresluts}, we can draw the following conclusions. First, when the number of measurements is relatively large (e.g., $N = 2400$), Tucker1 has the highest RMSE (poorest reconstruction) due to its limited degree of freedom. According to Eq.~\eqref{eq:tnntucker}, Tucker1 is mathematically equivalent to a Tucker decomposition model (defined in Eq.~\eqref{eq:tucker}) with a core tensor of dimension (5, 5, 5), which results in a larger representation error $E_1$ compared to the other two models. Second, although Tucker2 has slightly more parameters than TNN, it still has a larger RMSE due to its limited representation capability imposed by the multi-linear form. This demonstrates the importance of introducing non-linear activation functions in the model. Finally, the proposed TNN model achieves the best performance when the number of measurements is not too small, and shows a similar RMSE to Tucker1 under a small sampling ratio (e.g., $\rho=0.2$), thanks to its high expressive power and concise structure.

\subsubsection{Recoverability,  Trade-offs between being ``Deeper'' or ``Wider''}
To evaluate the recoverability and draw more insights into the proposed model, we compare the reconstruction performance of TNN models with different configurations (see Table~\ref{tab:TNN_params}) assuming no observation noise.  
Fig.~\ref{fig:tnns} shows the RMSEs under three different sampling ratios. More experimental results can be found in Appendix \ref{appendix-d}.

\begin{table}[t]
	\centering
	\caption{The configurations of TNN models. M is the number of parameters.}
	\label{tab:TNN_params}
	\begin{ruledtabular}
	\begin{tabular}{|c|c|c|c|}
	Name & L & Dimensionality & M \\
	\hline
	TNN1 & 3 & (5, 5, 5), (10, 10, 10), (20, 20, 20) & 875 \\
	\hline
	TNN2 & 2 & (10, 10, 10), (20, 20, 20) & 1600 \\
	\hline
	TNN3 & 2 & (15, 15, 15), (20, 20, 20) & 4275 
	\end{tabular}
	\end{ruledtabular}
\end{table}

\begin{figure}[htbp]
	\centering
	\includegraphics[width=1\reprintcolumnwidth]{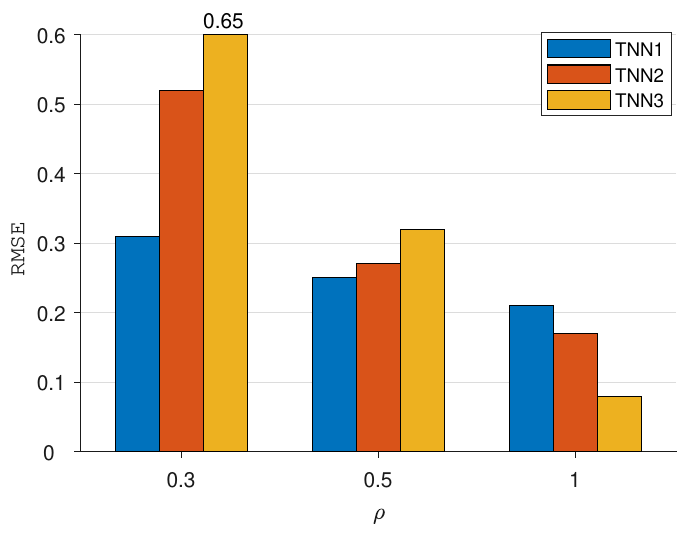}
	\caption{The averaged RMSEs of TNN models with different configurations under different sampling ratios.}
	\label{fig:tnns}
\end{figure}

It can be seen from Fig.~\ref{fig:tnns} that the recovery/reconstruction RMSEs of the proposed TNN models (with different layer numbers/configurations) keep decreasing to a very small number (e.g., $< 0.1$) as the sampling ratio increases (i.e., with more samples). This suggests that the recoverability of the proposed model is highly likely to be guaranteed when samples are abundant and clean.

On the other hand, despite having the largest number of layers, TNN1 gives the highest error when the $\rho=1$ (i.e., the SSF is fully observed). The reason is that TNN1 has the smallest number of model parameters among the three models, as shown in Table~\ref{tab:TNN_params}.
Generally, making the TNN ``deeper'' can significantly reduce the number of model parameters at the cost of slightly increasing the representation error. 
On the other hand, making the TNN ``wider'' can effectively increase expressiveness while also increasing the identification error, as seen in TNN3 in Table~\ref{tab:TNN_params} and Fig.~\ref{fig:tnns}.
The main idea behind the TNN model is to hierarchically reduce the tensor dimensionality through non-linear approximation, allowing for effective parameter reduction without significantly hampering expressiveness.

The experimental results indicate that designing the TNN model involves trade-offs between being wider or deeper. Although it is challenging to determine the optimal sweet spot theoretically, we can offer some practical suggestions on how to strike a good balance in practice:

\begin{itemize}
	\item It is advisable to keep the number of unknown model parameters less than the number of samples in order to potentially prevent ill-posed reconstruction problems.
	\item Consider increasing the dimensionality of the core tensor layer by layer when the number of layers is fixed. For example, the dimensionality of the $l$th layer can be set to twice that of the ($l-1$)th layer. This technique has shown promise in reducing the number of parameters while maintaining expressiveness.
	\item Recent studies have suggested that a one-layer Tucker model can effectively capture the main pattern of sound speeds, and adding just two or three more layers may be sufficient to capture non-linear variations and reduce the number of parameters. Consequently, the layer number of TNN is recommended to be set to 3 or 4.
\end{itemize}

It should be noted that these configurations were tested under the experimental settings outlined in Sec.~\ref{Sec IV-A}, and the optimal model configuration may vary depending on the specific application or scale.

\subsection{\label{sec:IV-C}Enhancement by TV Regularization}
\begin{figure}[!t]
    \center
    \includegraphics[width=1\reprintcolumnwidth]{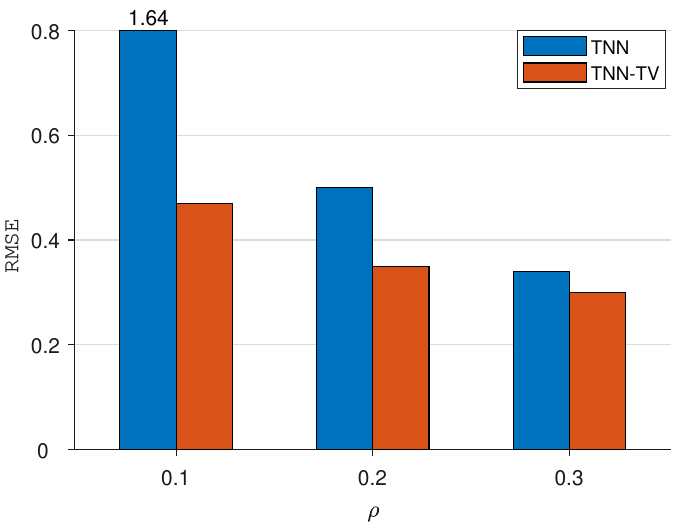}
    \caption{ The RMSEs of TNN-TV and TNN under different sampling ratios.}
    \label{fig:tnntv}
    \hrule
\end{figure}

In this subsection, experiments were conducted to demonstrate how the reconstruction performance of TNN can be further boosted by incorporating the structural information of the regularizer. For illustration purposes, the widely-used total variation (TV) \cite{rudin1992nonlinear} regularizer was used to enforce spatial smoothness among sound speeds. Specifically, the TV regularizer is defined as follows:
\begin{align}
    R(\bc{X})&=\sum_{i,j,k}(
    |\bc{X}(i+1,j,k)-\bc{X}(i,j,k)|+ 
    |\bc{X}(i,j+1,k) \\ \nonumber
    &-\bc{X}(i,j,k)|+
    |\bc{X}(i,j,k+1)-\bc{X}(i,j,k)|).
\end{align}
The reconstructed results of the proposed TNN model with TV regularizer, denoted as TNN-TV, are presented in Fig.~\ref{fig:tnntv}. The hyper-parameter $\lambda$ was manually tuned to optimize performance.

The results demonstrate that TNN-TV consistently outperforms TNN in SSF reconstruction, particularly under very sparse sampling (e.g., $\rho = 0.1$). Note that when the sampling ratio $\rho\geq 0.3$, the measurements can provide sufficient information for learning the model parameters, and hence TNN and TNN-TV have similar performances. In this case, $\lambda$ can be set to be nearly zero, and TNN-TV gradually reduces to TNN (see Eq.~\eqref{eq:tnnrec}). Since TNN is a special case of TNN-TV, in the next subsection, we only consider TNN-TV and refer to it as TNN for simplicity.

\subsection{Comparisons with SOTA methods}
Previous subsections have demonstrated the conciseness and expressiveness of the proposed model. In this subsection, we compare the reconstruction performance of the proposed algorithm with several SOTA SSF reconstruction methods.

 \textbf{SOTA Methods}: The selected SOTA methods include: 1) low-rank matrix-based method represented by the recently proposed graph-guided Bayesian matrix completion (BMCG);\cite{li2023graph} 2) tensor-based methods represented by alternating least squares for the Tucker model (Tucker-ALS), low-rank tensor completion (LRTC),\cite{zhang2014novel} and LRTC with total variation (LRTC-TV);\cite{jiang2018anisotropic} and 3) statistical learning-based methods represented by Gaussian process regression (GPR).\cite{rasmussen2004gaussian} 

\textbf{Model Settings}: The kernel of GPR is the widely used radial basis function (RBF), with the hyper-parameters being learned via evidence optimization.\cite{bishop2006pattern} In BMCG, the graph Laplacian matrix is constructed using the same kernel.\cite{li2023graph} The tensor rank surrogate function used in LRTC is the tensor nuclear norm, defined as $\sum_{k=1}^K\|\widetilde{\mathcal{X}}(:,:,k)\|_*$,\cite{zhang2014novel} where $\widetilde{\mathcal{X}}$ is the Fourier transformation of $\mathcal{X}$ along the third mode. 
\begin{table*}[!t]
    \begin{center}
	\caption{The average RMSEs over three Monte-Carlo trials of different algorithms under different sampling ratios and noise powers.}
	\label{tab:sota}
    \begin{ruledtabular}
        \begin{tabular}{|c|c|cccccc|}
        $\rho$ & $\sigma$ & \textbf{BMCG} & \textbf{GPR} & \textbf{Tucker-ALS} & \textbf{LRTC} & \textbf{LRTC-TV} & \textbf{TNN}\\ \hline
        
        \multirow{3}*{0.1} & 0.1 & 1.33 & 0.68 & 1.32 & 1.47 & 1.23 & \textbf{0.47} \\
        &0.3 & 1.34 & 0.71 & 1.36 & 1.50 & 1.25 & \textbf{0.51} \\
        &0.5 & 1.36 & 0.77 & 1.45 & 1.56 & 1.30 & \textbf{0.58}\\ \hline
        \multirow{3}*{0.2} & 0.1 & 0.72 & 0.52 & 0.76 & 0.96 & 0.77 & \textbf{0.35}\\
        &0.3 & 0.74 & 0.56 & 0.80 & 1.02 & 0.82 & \textbf{0.39}\\
        &0.5 & 0.77 & 0.61 & 0.84 & 1.12 & 0.89 & \textbf{0.46}\\ \hline
        \multirow{3}*{0.3} & 0.1 & 0.56 & 0.44 & 0.42 & 0.64 & 0.53 & \textbf{0.30}\\
        &0.3 & 0.59 & 0.48 & 0.44 & 0.73 & 0.60 & \textbf{0.33}\\
        &0.5 & 0.64 & 0.53 & 0.47 & 0.85 & 0.69 & \textbf{0.40}\\ \hline
        \multirow{3}*{0.4} & 0.1 & 0.48 & 0.37 & 0.41 & 0.46 &  0.39 & \textbf{0.28}\\
        & 0.3 & 0.52 & 0.44 & 0.42 & 0.57 & 0.48 & \textbf{0.31}\\
        & 0.5 & 0.57 & 0.49 & 0.45 & 0.71 & 0.59 & \textbf{0.34}
        \label{table:results}

        \end{tabular}
    \end{ruledtabular}	
    \end{center}
\end{table*}

\begin{figure*}[t]
    \center
    \includegraphics[width=1.8\reprintcolumnwidth]{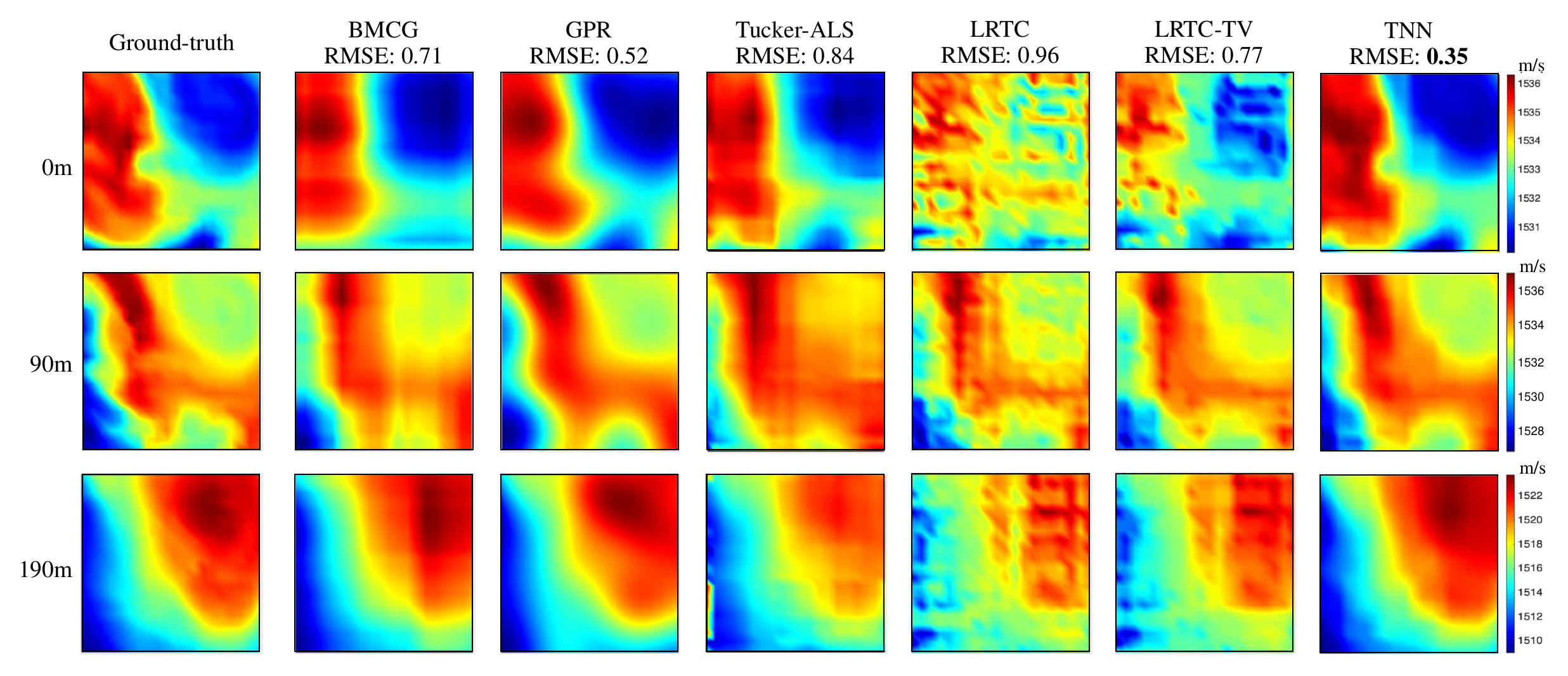}
    \caption{ Visual effects of the reconstructed SSF of different SOTA reconstruction methods in one single Monte-Carlo trial with sampling ratio $\rho=0.2$ and $\sigma=0.1$. The RMSEs are shown above the top subfigures.}
    \label{fig:sota20}
    \hrule
\end{figure*}

\begin{figure*}[t]
    \center
    \includegraphics[width=1.8\reprintcolumnwidth]{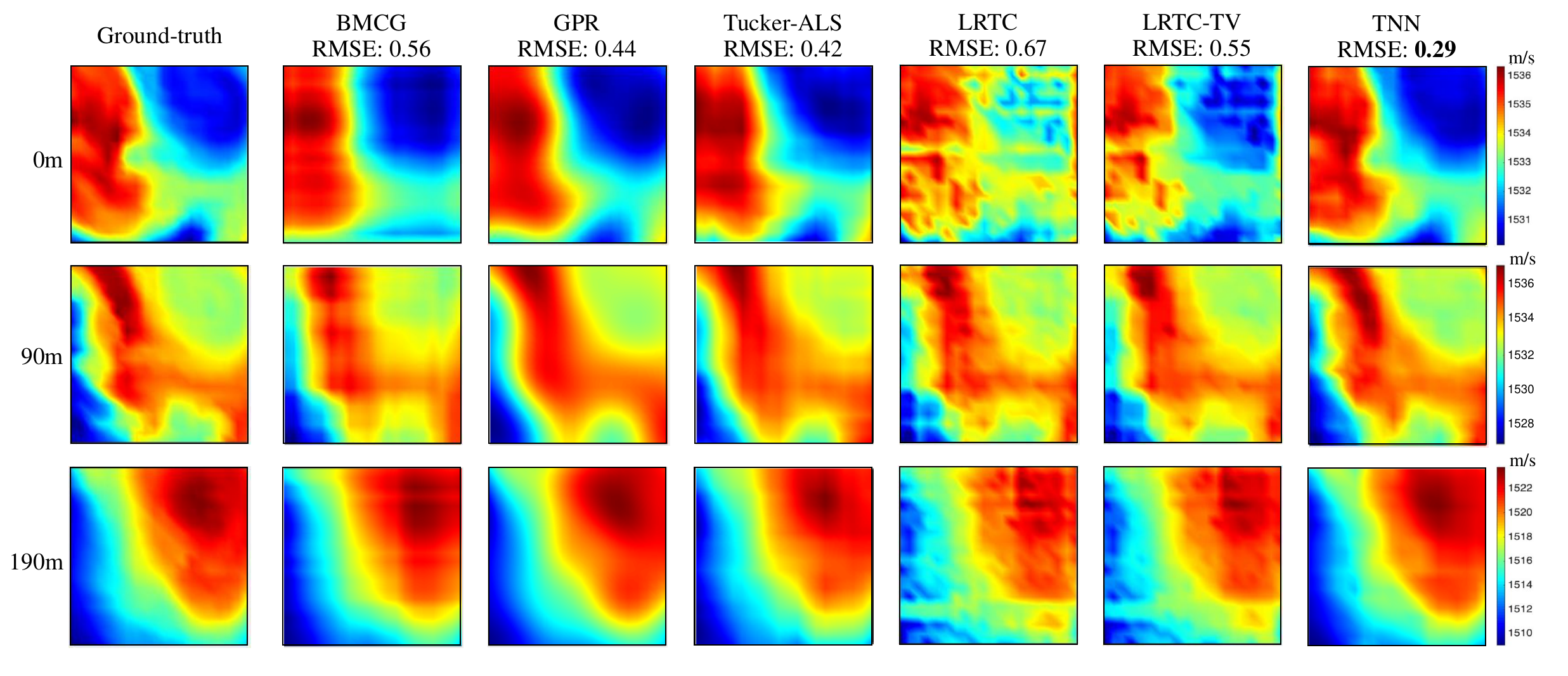}
    \caption{Visual effects of the reconstructed SSF of different SOTA reconstruction methods in one single Monte-Carlo trial with sampling ratio $\rho=0.3$ and $\sigma=0.1$. The RMSEs are shown above the top subfigures.}
    \label{fig:sota30}
    \hrule
\end{figure*}

\begin{figure*}[t]
    \center
    \includegraphics[width=1.8\reprintcolumnwidth]{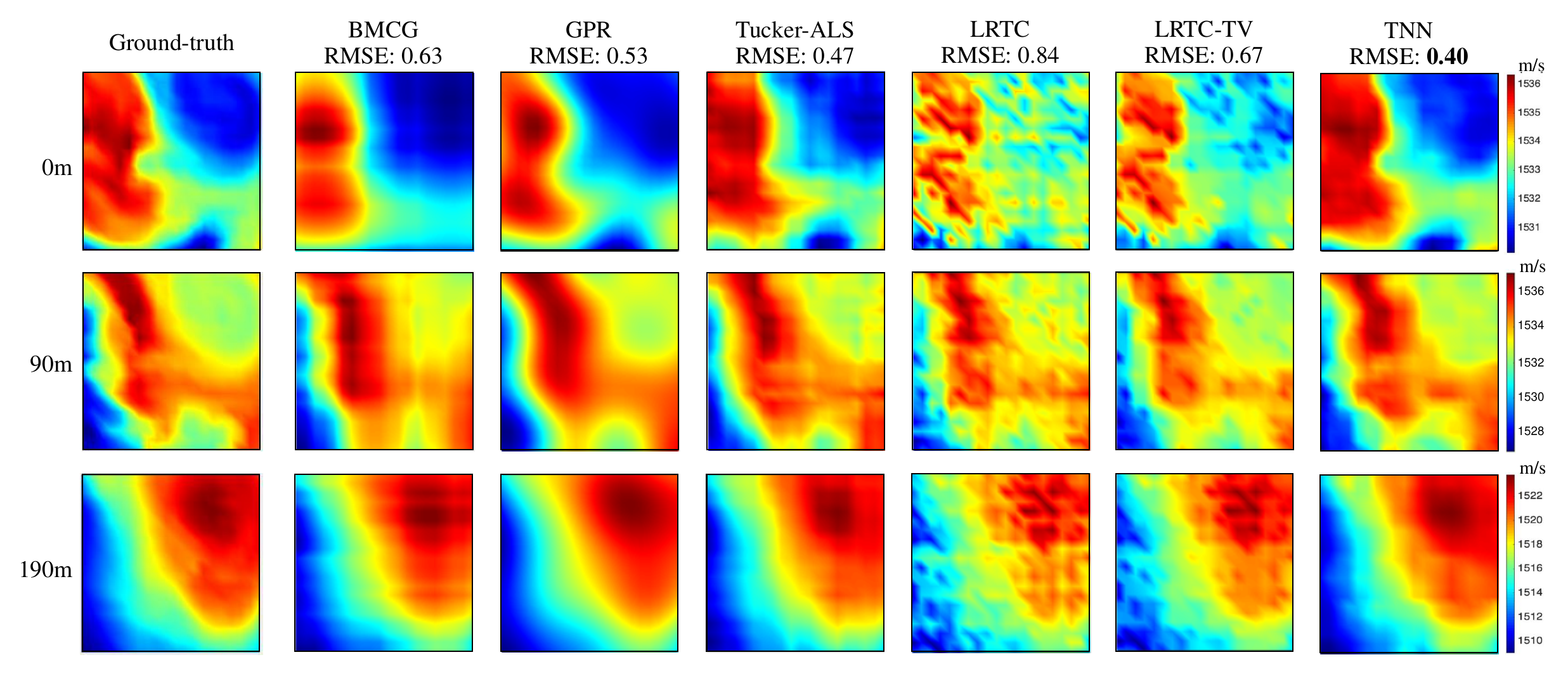}
    \caption{Visual effects of the reconstructed SSF of different SOTA reconstruction methods in one single Monte-Carlo trial with sampling ratio $\rho=0.3$ and $\sigma=0.5$. The RMSEs are shown above the top subfigures.}
    \label{fig:sota302}
    \hrule
\end{figure*}

\begin{figure*}[t]
    \center
    \includegraphics[width=1.8\reprintcolumnwidth]{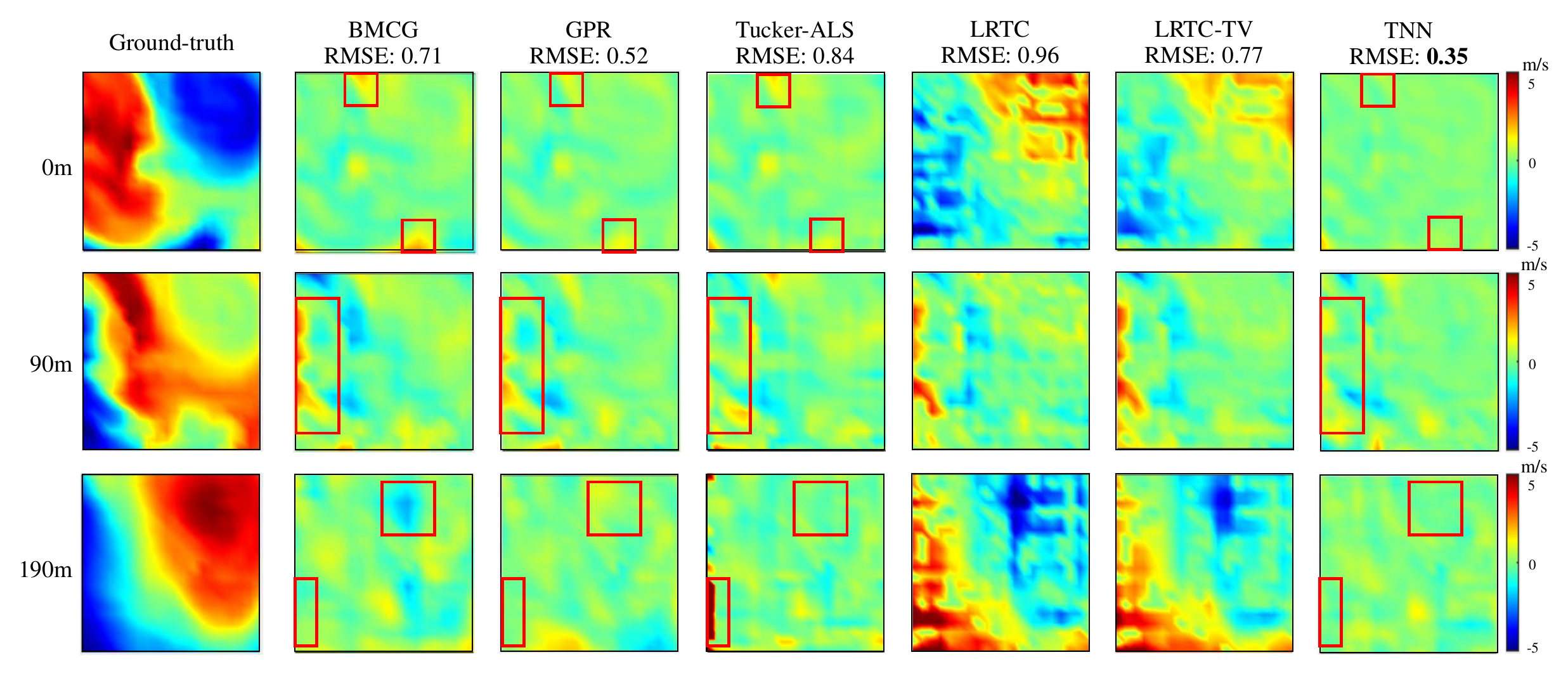}
    \caption{ Visual effects of the error surfaces in one single Monte-Carlo trial of different methods under $\rho=0.2$ and $\sigma=0.1$. The RMSEs are shown above the top subfigures.}
    \label{fig:errface}
    \hrule
\end{figure*}

\textbf{Results and Discussions}:  The reconstruction RMSEs of different methods under different sampling ratios and noise powers are shown in Table~\ref{tab:sota}.  Visual inspections of reconstructed SSFs and error surfaces under different scenarios are presented in Figs.~\ref{fig:sota20}-\ref{fig:errface}. As can be seen, the proposed tensor neural network-aided reconstruction algorithm outperforms all other methods in terms of reconstruction accuracy, regardless of the sampling ratio and noise power. The reason is that the proposed model has a high expressive power with relatively few parameters, achieving an outstanding trade-off between $E_1$ and $E_2$, as discussed in Sec.~\ref{sec:3}. Additionally, the TNN model exhibits strong noise rejection (as proved in \textbf{Proposition 1}), making it robust against measurement noise. In the following, we provide additional discussions to draw further insights from the results.

\begin{figure}[!t]
    \center
    \includegraphics[width=1\reprintcolumnwidth]{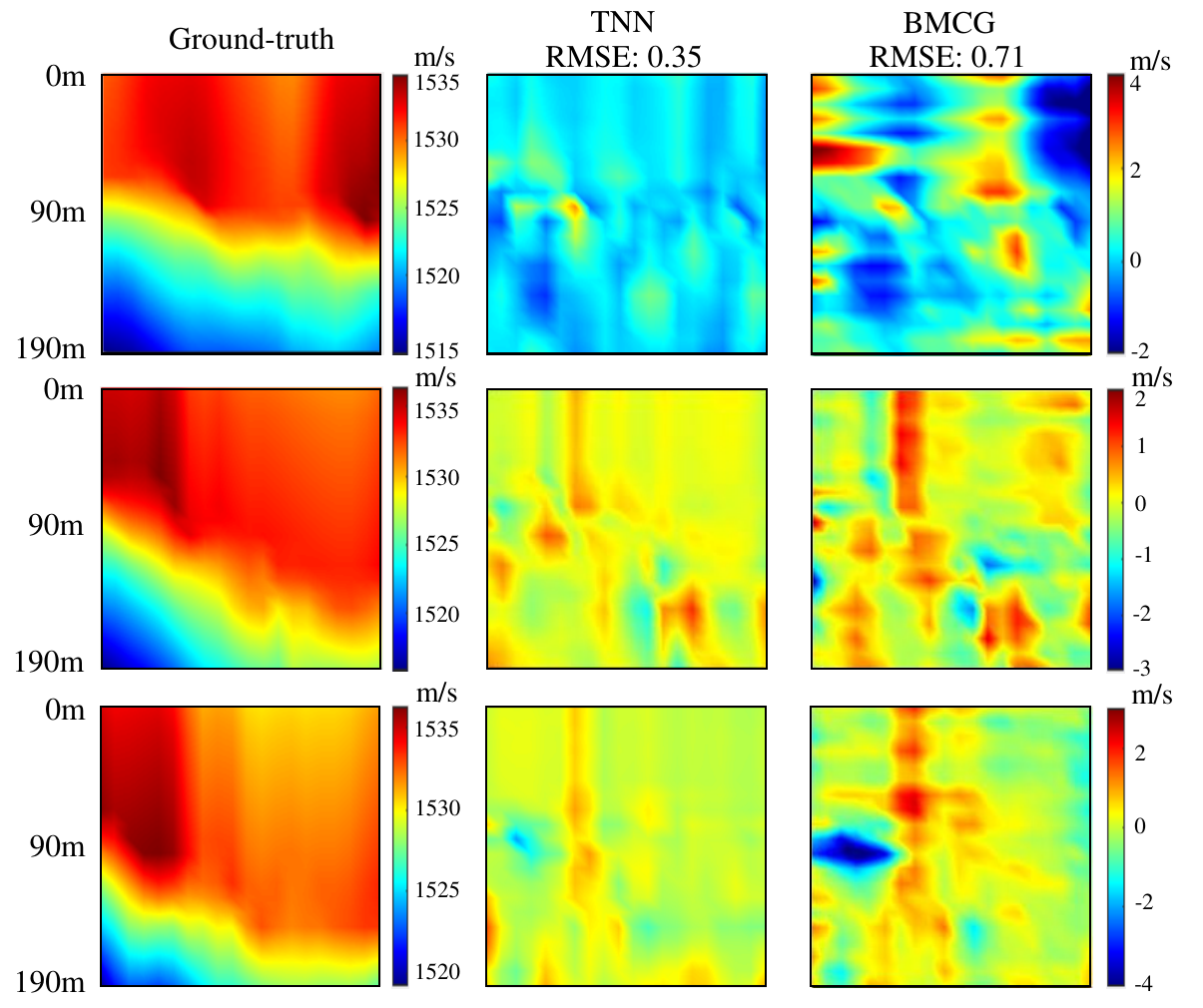}
    \caption{ The error surfaces of tensor-based method TNN and matrix-based method BMCG in the vertical dimension in one Monte-Carlo trial. The TNN can exploit multi-dimensional correlations thus having smoother error surfaces.}
    \label{fig:vertical}
    \hrule
\end{figure}

\begin{itemize}
    \item The matrix-based method, BMCG, yields good reconstruction results for the horizontal SSF slices by explicitly exploring the spatial smoothness through graph models, as discussed in Sec. III of Ref.~\citen{li2023graph}. However, the vertical correlations of sound speeds are ignored in this method, leading to the loss of fine-grained details in the vertical dimension. This can be observed in the error surfaces of the vertical dimension shown in Fig.~\ref{fig:vertical}.
    
    \item Among tensor completion methods (i.e., Tucker-ALS, LRTC, and LRTC-TV), Tucker-ALS gives the best reconstruction performance. It outperforms the matrix-based method BMCG when the observation is moderate (e.g., $\rho = 0.3$) since it can effectively exploit the multi-dimensional correlations.
    	However, the multi-linear form has limited the expressive power of Tucker-ALS, resulting in a considerable $E_1$, making the associated reconstruction performance inferior to the proposed one.  Moreover, these tensor completion methods are sensitive to the sampling pattern and exhibit significantly degraded performance when the measurements are sparse.
    
    \item GPR is a non-parametric statistical model that paves another promising path for reconstructing SSFs. The expressiveness of the kernel determines the representation error $E_1$, while the learning process of the kernel hyper-parameters determines the identification error $E_2$. Although GPR has demonstrated great success in many areas in recent years,\cite{caviedes2021gaussian,michalopoulou2021matched,michalopoulou2022uncertainty} research on its applications in ocean SSF reconstruction is still in its infancy (including optimal kernel design and hyper-parameter learning). In our experiments, GPR, using the RBF kernel and evidence maximization-based hyper-parameter learning, achieved the second-best performance in almost all scenarios. However, to enhance its performance, more powerful kernels and advanced optimization techniques can be employed,\cite{li2023graph,calandriello2022scaling} which require significant research efforts and are beyond the scope of this paper. Moreover, the high computational complexity of GPR limits its widespread application. The scalability of GPR is an active research area in machine learning, as evidenced by recent works.\cite{xu2020exact,ober2021global}
   
\end{itemize}

\section{\label{sec:5} Conclusions and Future Directions}
In this paper, under the unified framework of analyzing the reconstruction error, a tensor neural network-aided reconstruction algorithm is proposed. By leveraging the succinct form of tensor models and the high expressive power of deep neural networks, the proposed TNN model can concisely and accurately represent the 3D SSF, achieving an outstanding balance between the representation error $E_1$ and identification error $E_2$. A simple and efficient algorithm is developed based on the widely adopted gradient descent method. Experiments using real-life 3D SSF data demonstrate the effectiveness of the proposed reconstruction error analysis framework and showcase the encouraging performance of the proposed algorithm compared to other SOTA reconstruction methods. Our study demonstrated how the proposed model could effectively integrate structural assumptions, as exemplified by the use of the TV regularizer.

Future research could explore the incorporation of additional prior information, such as the physical knowledge of ocean processes that drive sound speed variations, to achieve even better reconstruction performance. It would also be valuable to extend the analysis in Ref.~\citen{shrestha2022deep} to quantify the recoverability of the proposed TNN model and establish the theoretical connection between TNN and CNN with 3D convolutions. Finally, rethinking the reconstruction problem from a Bayesian learning perspective \cite{cheng2022rethinking,cheng2022towards} would hold significant research value. This approach allows for automatic control of model complexity to prevent both overfitting and underfitting, and enables the utilization of uncertainty information to guide future sensing planning.

\section{\label{sec:6} acknowledgement}
This work was supported in part by the  National Key R\&D Program of China, in part by the National Natural Science Foundation of China under Grant 62001309, in part by the Fundamental Research Funds for the Central Universities, in part by the Zhejiang University Education Foundation Qizhen Scholar Foundation, and in part by Science and Technology on Sonar Laboratory under Grant 6142109KF212204.

\appendix
\section{tensor operation definitions}
\label{appendix-a}
{\it Definition 1 (mode-n product)}: The mode-$n$ product of a tensor $\bc{A}\in\mathbb{R}^{I_1\times\cdots\times I_N}$ and a matrix $\mathbf{B}\in\mathbb{R}^{J\times I_n}$ is defined by
\begin{align}
    \bc{C}=\bc{A}\times_{n}\mathbf{B}\in\mathbb{R}^{I_1\times\cdots\times I_{n-1}\times J\times I_{n+1}\times\cdots\times I_N},
\end{align}
whose entries are given by
\begin{align}
    \bc{C}(i_1,\cdots,j,\cdots,i_N)=\sum_{i_n=1}^{I_n}\bc{A}(i_1,\cdots,i_n,\cdots,i_N)\mathbf{B}(j,i_n).
\end{align}

{\it Definition 2 (mode-n unfolding)}: Given a tensor $\bc{A}\in\mathbb{R}^{I_1\times\cdots\times I_N}$, its mode-$n$ unfolding gives a matrix $\mathbf{A}_{(n)}\in \mathbb{R}^{I_n\times\prod_{k=1,k\neq n}^NI_k}$. Each tensor element $\bc{A}(i_1,\cdots,i_n)$ is mapped to the matrix element $[\mathbf{A}_{(n)}](i_n,j)$ where $j=1+\sum_{k=1,k\neq n}^N(i_k-1)J_k$ with $J_k=\prod_{m=1,m\neq n}^{k-1}I_m$. 

{\it Definition 3 (Tucker decomposition)}: For a tensor $\bc{A}\in\mathbb{C}^{I_1\times\cdots\times I_N}$, the Tucker decomposition is defined as
\begin{align}
\bc{A}=\bc{G}\times_1\mathbf{U}^{(1)}\times_2\mathbf{U}^{(2)}\times_3\cdots\times_N\mathbf{U}^{(N)},
\end{align}
where each factor matrix $\mathbf{U}^{(n)}\in \mathbb{C}^{I_n\times R_n},\forall n=1,\cdots,N$. The core tensor $\bc{G}\in\mathbb{C}^{R_1\times\cdots\times R_N}$. The tuple $(R_1,\cdots, R_N)$ is known as the multi-linear rank.

{\it Definition 4 (tensor inner product)}: The inner product of tensor $\bc{A}$ and $\bc{B}$ with the same size is formulated as follows:
\begin{align}
    c=\langle\bc{A},\bc{B}\rangle=\langle \text{vec}(\bc{A}), \text{vec}(\bc{B}) \rangle \in \mathbb{R},
\end{align}
where $\text{vec}(\bc{A})$ is the vectorization of $\bc{A}$.

\section{proof and derivation}
\label{appendix-b}
\subsection{Derivation of Eq.~\eqref{eq:ed}}
According to its definition, the reconstruction error $E$ can be written as
\begin{align}
    E&=\|\bc{X}-\hat{\bc{X}}\|_\F^2  \\ \nonumber
&=\|\bc{X}-\bc{D}(\hat{\bs{\Theta}})\|_\F^2 \\ \nonumber
&=\|\bc{X}-\bc{D}(\bs\Theta^*)+\bc{D}(\bs\Theta^*)-\bc{D}(\hat{\bs{\Theta}})\|_\F^2 \\ \nonumber
&=\underbrace{\|\bc{X}-\bc{D}(\bs\Theta^*)\|_\F^2}_{\triangleq~ E_1}+\underbrace{\|\bc{D}(\bs\Theta^*)-\bc{D}(\hat{\bs{\Theta}})\|_\F^2}_{\triangleq~E_2} + \\ \nonumber
&~~~~~\underbrace{2\langle \bc{X}-\bc D( {\bs \Theta}^*), \bc D( {\bs \Theta}^*)-\bc{D}(\hat{\bs \Theta})\rangle}_{\triangleq~ \epsilon}.
\end{align}
Thus the reconstruction error can be decomposed as $E=E_1+E_2+\epsilon$ and Eq.~\eqref{eq:ed} holds.

\subsection{Derivation of Eq.~\eqref{eq:cross}}
If $\bc{X}=\bc{D}(\bs\Theta^*)$, we have $E_1=\|\bc{X}-\bc{D}(\bs\Theta^*)\|_\F^2=0$ and $\epsilon=2\langle \mathbf{0}, \bc D( {\bs \Theta}^*)-\bc D(\hat{\bs \Theta})\rangle=0$. Similarly, if $\bc{D}(\bs{\Theta}^*)=\bc{D}(\hat{\bs{\Theta}})$, then $E_2=\|\bc{D}(\bs{\Theta}^*)-\bc{D}(\hat{\bs{\Theta}})\|_\F^2=0$ and $\epsilon=2\langle \bc{X}-\bc D( {\bs{\Theta}}^*),\mathbf{0}\rangle=0$.

\subsection{Proof of Proposition 1}
We start by rewriting the term $\|\bc{X}-\bc{E}\|_\F^2$ in a convenient form. Specifically, since each element of $\bc{E}$ follows a zero-mean i.i.d Gaussian distribution, we have $\|\bc{X}-\bc{E}\|_\F^2= \|\text{vec}(\bc{X})-\zeta\|_\F^2$, where $\text{vec}(\bc{X}) \in \mathbb{R}^T$ is the vectorization of $\bc{X}$ and $\zeta \sim \mathcal{N}(\mathbf{0},\sigma^2\mathbf{I})$ where $\mathbf{I}$ is a $T\times T$ identity matrix. Consider a set of parameters $\mathbf{\Theta}=\{\bc{G},\mathbf{W}^{(1)}_1,\mathbf{W}^{(2)}_1,\mathbf{W}^{(3)}_1\}$, denote $\bc{O}$ as the indicator tensor that contains one if $\bc{X}_{i,j,k}>0$ and zero otherwise. Then we have
\begin{align}
    \text{vec}(\bc{X})&=\text{vec}\left(\text{ReLU}(\bc{G}\times_1 \mathbf{W}^{(1)}_1 \times_2 \mathbf{W}^{(2)}_1 \times_3 \mathbf{W}^{(3)}_1)\right) \\ \nonumber
    &=\text{vec}\left(\bc{O}*(\bc{G}\times_1 \mathbf{W}^{(1)}_1 \times_2 \mathbf{W}^{(2)}_1 \times_3 \mathbf{W}^{(3)}_1)\right) \\ \nonumber
    &=\underbrace{\left[\text{diag}(\text{vec}(\bc{O}))(\mathbf{W}^{(3)}_1\otimes \mathbf{W}^{(2)}_1\otimes\mathbf{W}^{(1)}_1 )\right]}_{\triangleq~\mathbf{B}\in \mathbb{R}^{T\times R}}\text{vec}(\bc{G}) \\ \nonumber
    &=\mathbf{B}\text{vec}(\bc{G}).
\end{align}
Therefore, $\text{vec}(\bc{X})$ is a weighted combination of $R$ $T$-dimensional vectors and lies in a at-most-$R$-dimensional subspace $S$ of $\mathbb{R}^{T}$. It follows that 
\begin{align}
    \min_{\bs\Theta}\|\bc{X}-\bc{E}\|_\F^2\geq \|P_{S^c\zeta}\|_2^2, \label{eq:ieqnoise}
\end{align}
where $P_{S^c\zeta}$ is the projection of $\zeta$ outside the subspace $S$.

Next, we make use of the following lemma to give a bound on the projection of the noise $\zeta$ onto a subspace.

 \begin{tcolorbox}
\textbf{Lemma 1}: Let $S\subset \mathbb{R}^T$ be a subspace with dimension $R$. Let $\zeta\sim\mathcal{N}(\mathbf{0},\sigma^2\mathbf{I})$ and $\beta\geq 1$. Then,
\begin{align}
    P\left[ \frac{\|P_{S^c\zeta}\|_2^2}{\|\zeta\|_2^2}\geq 1-\frac{10\beta R}{T} \right]\geq 1-e^{-\beta R}-e^{-T/16}.
\end{align} \\
{\it Proof}: From the Lemma 1 in Ref.~\citen{laurent2000adaptive}, if $X\sim \chi_T^2$, then
\begin{align}
    P[X-T\geq 2\sqrt{Tx}+2x]\leq e^{-x}, \label{eq:ineq1}\\ 
    P[X\leq T-2\sqrt{Tx}]\leq e^{-x} \label{eq:ineq2}.
\end{align}
We have $\frac{\|P_{S^c\zeta}\|_2^2}{\|\zeta\|_2^2}=1-\frac{\|P_{S\zeta}\|_2^2}{\|\zeta\|_2^2}$. Note that $\|P_{S\zeta}\|_2\sim \chi_R^2$ and $\|\zeta\|_2^2\sim \chi_T^2$. Applying the inequality \eqref{eq:ineq1} to bound $\|P_{S\zeta}\|_2$ and inequality \eqref{eq:ineq2} to $\|\zeta\|_\F^2$, a union bound gives that claim.
\end{tcolorbox}
With Lemma 1 and inequality \eqref{eq:ieqnoise}, we have 
\begin{align}
    P\left[\frac{1}{\|\bc{E}\|_\F^2}(\min_{\bs\Theta}\|\bc{X}-\bc{E}\|_\F^2) \geq 1-\frac{10\beta R}{T}\right]\geq 1-e^{-\beta R}-e^{-T/16},
\end{align}
for $\beta\geq 1$. Let $\beta=1$, then
\begin{align}
    P\left[ \min_{\bs\Theta}\|\bc{X}-\bc{E}\|_\F^2 \geq \|\bc{E}\|_\F^2(1-\frac{10R}{T}) \right]\geq 1-e^{-R}-e^{-T/16}.
\end{align}
Thus Proposition 1 holds.


\subsection{Derivation of Eq.~\eqref{eq:tnntucker}}
Before giving the derivation of Eq.~\eqref{eq:tnntucker}, we present a few facts regarding $n$-mode matrix products. Specifically, for distinct modes in a series of multiplications, the order of the multiplication is irrelevant, i.e.,
\begin{align}
    \bc{X}\times_m \mathbf{A} \times_n \mathbf{B} = \bc{X}\times_n \mathbf{B}\times_m \mathbf{A}~~ (m\neq n).
\end{align}
If the modes are the same, then 
\begin{align}
    \bc{X}\times_m \mathbf{A} \times_m \mathbf{B} = \bc{X}\times_m (\mathbf{B}\mathbf{A}).
\end{align}
Given these facts, a $L$-layer TNN with a linear activation function can be written as
\begin{align}
    \bc{X}&=\varsigma(\cdots\varsigma\left( \bc{G}\times_1 \mathbf{W}^{(1)}_1 \times_2 \mathbf{W}^{(2)}_1 \times_3 \mathbf{W}^{(3)}_1  \right) \cdots \\ \nonumber
    &~~~~\times_1 \mathbf{W}^{(1)}_L \times_2 \mathbf{W}^{(2)}_L \times_3 \mathbf{W}^{(3)}_L) \\ \nonumber
    &=a(\cdots a\left( \bc{G}\times_1 \mathbf{W}^{(1)}_1 \times_2 \mathbf{W}^{(2)}_1 \times_3 \mathbf{W}^{(3)}_1  \right) \cdots \\ \nonumber
    &~~~~\times_1 \mathbf{W}^{(1)}_L \times_2 \mathbf{W}^{(2)}_L \times_3 \mathbf{W}^{(3)}_L) \\ \nonumber
    &= a^L\bc{G} \times_1 \mathbf{W}^{(1)}_1 \cdots \times_1 \mathbf{W}^{(1)}_L  \times_2 \mathbf{W}^{(2)}_1 \cdots  \\ \nonumber
    &~~~~\times_2 \mathbf{W}^{(2)}_L \times_3 \mathbf{W}^{(3)}_1 \cdots \times_3 \mathbf{W}^{(3)}_L \\ \nonumber
    &= a^L\bc{G}\times_1(\mathbf{W}^{(1)}_L\cdots\mathbf{W}^{(1)}_1) \times_2(\mathbf{W}^{(2)}_L\cdots\mathbf{W}^{(2)}_1)\\ \nonumber &~~~~\times_3(\mathbf{W}^{(3)}_L\cdots\mathbf{W}^{(3)}_1) \\ \nonumber
    &=\bar{\bc{G}}\times_1 \bar{\mathbf{W}}^{(1)} \times_2 \bar{\mathbf{W}}^{(2)} \times_3 \bar{\mathbf{W}}^{(3)}.
\end{align}

\section{derivation of the gradient}
\label{appendix-c}
The gradient with respect to the model parameters can be obtained via the backpropagation technique.\cite{rojas1996backpropagation} 
Since the gradient of the activation function depends on the specific choice, here we mainly introduce the gradient propagation in the TCL. By unfolding the tensor to the matrix, we can calculate the gradients with respect to the factor matrices, given by
\begin{align}
    \frac{\partial \mathbf{X}^{l+1}_{(k)}}{\partial \mathbf{W}^{(k)}_l}=\frac{\partial \mathbf{W}^{(k)}_l \mathbf{X}^l_{(k)}(\mathbf{W}^{(-k)}_l)^\T}{\partial \mathbf{W}^{(k)}_l}
\end{align}
where $\mathbf{W}^{(-k)}_l=\mathbf{W}^{(3)}_l\otimes\cdots\mathbf{W}^{(k+1)}_l\otimes\mathbf{W}^{(k-1)}_l\cdots\mathbf{W}^{(1)}_l$, $k=1,2,3$;  $\mathbf{X}^l_{(k)}$ is the mode-$k$ unfolding of $\bc{X}_{l}$ and $\mathbf{X}^{l+1}_{(k)}$ is the mode-$k$ unfolding of $\bc{X}_{l+1}$. Similarly, the gradient with respect to the input tensor can be obtained by vectorizing the tensor, given by
\begin{align}
    \frac{\partial \text{vec}(\bc{X}_{l+1})}{\partial \text{vec}(\bc{X}_{l})}=\frac{\partial (\mathbf{W}^{(3)}_l\otimes\mathbf{W}^{(2)}_l\otimes\mathbf{W}^{(1)}_l)\text{vec}(\bc{X}_{l})}{\partial \text{vec}(\bc{X}_{l})}.
\end{align}
Then the gradient of the loss function with respect to the parameters can be calculated by the chain rule.

\section{further experimental results}
\label{appendix-d}
Table~\ref{tab: append} shows the RMSEs of various TNN models at different sampling ratios. The model configurations used are the same as in Table~\ref{tab:TNN_params}.
\begin{table}[t]
	\centering
	\caption{The RMSEs of different TNN models under different sampling ratios.}
	\label{tab: append}
	\begin{ruledtabular}
	\begin{tabular}{|c|c|c|c|c|c|c|c|}
	\diagbox{Name}{$\rho$} & 0.1 & 0.2 & 0.4 & 0.6 & 0.7 & 0.8 & 0.9\\
		\hline
		TNN1 & \textbf{1.64} & \textbf{0.55} & \textbf{0.27} & 0.23 & 0.22 & 0.21 & 0.21\\
		\hline
		TNN2 & 4.10 & 1.00 & 0.30 & \textbf{0.21} & \textbf{0.19} & 0.18 & 0.17\\
		\hline
		TNN3 & 4.66 & 2.03 & 0.40 & 0.24 & 0.21 & \textbf{0.15} & \textbf{0.09}
		\end{tabular}
	\end{ruledtabular}
\end{table}

\bibliography{sampbib}

\end{document}